\documentclass[twocolumn]{aastex631}

\hypersetup{linkcolor=blue,citecolor=blue,filecolor=cyan,urlcolor=Maroon}
\usepackage{amsmath}
\usepackage{txfonts}
\usepackage[dvipsnames]{xcolor}
\usepackage{longtable}
\usepackage{threeparttable}
\usepackage{multirow}
\usepackage{comment}
\shorttitle{Hubble flow around M~81}
\shortauthors{J.~Wagner, D.~Benisty, and I.~Karachentsev}
\graphicspath{{./}{}}

\begin{document}

\title{\Large{\textnormal{\textsc{The Binary Ballet: Mapping The Local Expansion Around M~81 \& M~82}}}}

\author[0000-0002-4999-3838]{Jenny Wagner}
\affiliation{Helsinki Institute of Physics, P.O. Box 64, FI-00014 University of Helsinki, Finland, \textcolor{Maroon}{{wagner@asiaa.sinica.edu.tw}}}
\affiliation{Academia Sinica Institute of Astronomy and Astrophysics, 11F of AS/NTU Astronomy-Mathematics Building, Roosevelt Rd, Taipei 106216, Taiwan, R.O.C}
\affiliation{Bahamas Advanced Study Institute and Conferences, 4A Ocean Heights, Hill View Circle, Stella Maris, Long Island, The Bahamas}

\author[0000-0002-9578-3081]{David Benisty}
\affiliation{Leibniz-Institut f\"ur Astrophysik Potsdam (AIP), An der Sternwarte 16, 14482 Potsdam, Germany, \textcolor{Maroon}{benidav@aip.de}}

\author[0000-0003-0307-4366]{Igor D.~Karachentsev}
\affiliation{Special Astrophysical Observatory, The Russian Academy of Sciences, Nizhnĳ Arkhyz, Karachai-Cherkessian Republic 369167, Russia \textcolor{Maroon}{idkarach@gmail.com}}

\begin{abstract} 
This study of the M~81 complex and its Hubble flow delivers new and improved Tip of the Red Giant Branch (TRGB)-based distances for nine member galaxies, yielding a total of 58 galaxies with high-precision TRGB distances. With those, we perform a systematic analysis of the group's dynamics in the core and its embedding in the local cosmic environment. 
Our analysis confirms that the satellite galaxies of the M~81 complex exhibit a flattened, planar distribution almost perpendicular to the supergalactic pole and thus aligned with a larger-scale filamentary structure in the Local Universe.
We demonstrate that the properties of the group's barycentre are robustly constrained by the two brightest members, M~81 and M~82, and that correcting heliocentric velocities for the solar motion in the Local Group decreases the velocity dispersion of the group. 
Then applying minor and major infall models, we fit the local Hubble flow to constrain the Hubble Constant and the total mass of the M~81 complex. The joint best-fit parameters from both models yield $H_0 = \left(63 \pm 6 \right)$~km/s/Mpc and total mass of $(2.28\pm 0.49) \times 10^{12}~M_{\odot}$. We thus arrive at an increased mass estimate compared to prior work but reach a higher consistency with virial, $(2.74 \pm 0.36)\times 10^{12}\,M_\odot$, and projected-mass estimates, $(3.11 \pm 0.69)\times 10^{12}~M_\odot$. Moreover, our $H_0$ estimate shows an agreement with Planck, consistent with other TRGB-based Local-Universe inferences of $H_0$ and still within a 2-$\sigma$ agreement with Cepheid-based Local-Universe probes.
\end{abstract}


%
\nolinenumbers  
\section{Introduction}
\label{sec:introduction}
The M~81 galaxy complex stands as one of the most prominent and massive groups in the nearby Universe, providing a critical laboratory for studying galaxy interactions, group dynamics, and the small-scale properties of the Hubble flow~\citep{bib:Karachentsev2006,bib:Karachentsev2021}. 
Our study introduces previously unmeasured and improved TRGB-based distances for nine galaxies: M~82, d0944+71, KDG 73, UGC 4483, DDO 53, dw0959+68, HolmII, KKH37, and UGC 6456. Among these, the distance to M~82 is particularly vital for dynamically determining the group's centre of mass, while galaxies like HolmII and UGC 6456, located at distances exceeding 1 Mpc from M81, provide crucial anchors in the Hubble flow regime. This comparably homogeneous TRGB-based sample is supplemented by seven additional galaxies (JKB83, KDG61em, Clump I, Clump III, HIJASS1021+68, UGC 6451, and SMDG0956+82) with distances newly inferred from other methods; the first six are based on associations with neighbouring members, while the last, SMDG0956+82, relies on the numerical action method (NAM) and is consequently too uncertain for our primary analysis despite its location beyond 1 Mpc.
Moreover, the first four are emission sparks that do not contain old stellar populations and HIJASS1021+68 is a starless HI cloud, such that the TRGB-approach cannot be applied to these galaxies.

With this extended and improved data set based on the Catalog and Atlas of Local Volume Galaxies (see \cite{bib:Karachentsev2013} and references therein), we perform a systematic analysis to determine the position and line-of-sight velocity of the group's centre of mass employing the four brightest central galaxies (M~81, M~82, NGC~3077, and NGC~2976), finding that M~81 and M~82 alone are sufficient to robustly constrain these properties. 
We then use all 58 galaxies with precise TRGB-based distances to constrain the Hubble Constant, $H_0$, and the total enclosed mass of the M~81 complex, $M$, through a detailed Hubble flow analysis. 
This includes a thorough investigation of the impact of key nuisance parameters on the final inferred cosmological and dynamical parameters.
Most prominently, these nuisance parameters are the correction of the solar motion in the Local Group and the chosen starting and ending distances for the Hubble flow fit.  By our systematic analyses of the local expansion field using all galaxies with TRGB-based distances available, this study provides new, precise benchmarks for testing cosmological models on small scales and for understanding the mass assembly of a major galaxy group in a filamentary embedding.

The paper is organised as follows: Sect.~\ref{sec:data_set} describes the characteristics of the data set and compares it to the one previously used in \cite{bib:Mueller2024} and \cite{bib:Wagner2025}. 
Sect.~\ref{sec:embedding_environment} discusses the embedding of the M~81 complex into the local cosmic environment. 
Sect.~\ref{sec:galaxy_infall} presents infall models to estimate the radial velocities of galaxies relative to the group's barycentre and also details the preprocessing of the measurements to apply these models, in particular the bias correction to the heliocentric velocities and the determination of the group's barycentre. 
Then, Sect.~\ref{sec:Hubble_flow} details the Hubble flow fitting function and approaches to constrain $H_0$ and $M$ from our data. 
Sect.~\ref{sec:dynamical_mass} compares the mass estimates from the Hubble flow fit with complementary dynamical mass estimators. 
Finally, Sect.~\ref{sec:discussion} summarises our results and puts them into the context of prior works and the role of Local-Universe Hubble flow fits for the $H_0$ tension. 


\section{Data set}
\label{sec:data_set}
\noindent
The total data set contains 98 galaxies around M~81 and in its embedding environment. 
Of those, 26 do not have velocity measurements and are excluded from the analysis.
Out of the remaining 72 galaxies, 58 have TRGB-based distances, 6 distances were obtained by applying the numerical action method to galaxies having TRGB-based distances and inferring the distance from the reconstructed dynamical field, as detailed in \cite{bib:Shaya2017} and \cite{bib:Kourkchi2020}, 6 distances were obtained from the galaxy membership in known groups with measured distances of other members (mem), one distance was obtained by surface brightness fluctuations (SBF), and one distance was obtained based on the texture (txt), i.~e.~the granular structure of the galaxy, see also \cite{bib:Karachentsev2013} for details. 
All methods are subject to various calibrations and physical mechanisms, so that their observational uncertainties differ. 
We assume 5\% uncertainties on the TRGB-based distances and estimate the other methods to be 25\%-30\% imprecise. 

\begin{figure}
\centering
\includegraphics[width=0.98\linewidth]{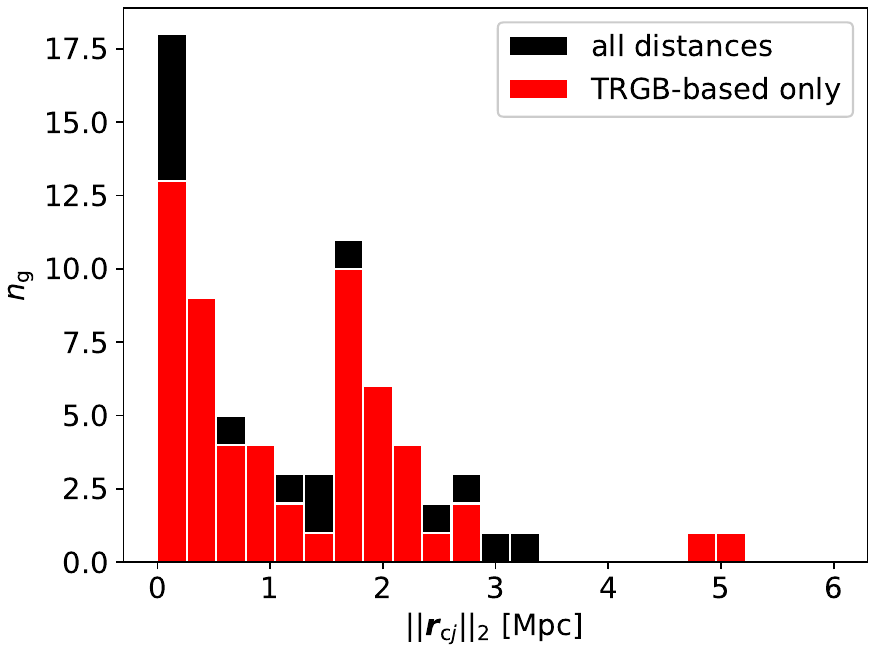}
\caption{\it{Histogram of distances for all 71 galaxies with respect to M~81 as the centre (black bars) and only those 57 galaxies having TRGB-based distances (red bars). The bin size is 0.25~Mpc.}}
\label{fig:hist_dist}
\end{figure}

Fig.~\ref{fig:hist_dist} shows a histogram of distances around M~81 for all distance measures (black bars) and TRGB-based distances only (red bars) using a binning of 0.25~Mpc. 
As can be read off this plot, the 58 galaxies having TRGB-based distances are sufficient to perform a Hubble flow analysis and we can exclude galaxies with less precise distance measurements from the analysis. 

These 58 galaxies with TRGB-based distances were then grouped by finding their ``major disturber" (MD), meaning their closest neighbouring galaxy that exerts the maximum tidal impact on the galaxy as detailed in \cite{bib:Karachentsev1999}. 
The ``tidal index", $\Theta_1$, for a galaxy~$j$ to quantify this impact and determine the MD is given by 
\begin{equation}
\Theta_1 \equiv \mathrm{max} \left\{ \log\left(\frac{M_k}{\Vert r_{jk} \Vert_2} \right) \right\} + C \,, \quad \forall k \ne j \,,
\label{eq:theta1}
\end{equation}
in which $M_k$ denotes the mass of the $k$-th neighbouring galaxy and $\Vert r_{jk} \Vert_2$ denotes the three-dimensional distance between galaxy $j$ and $k$.
The constant $C$ is chosen such that a $\Theta_1 = 0$ implies that galaxy~$j$ lies exactly at the zero-velocity surface of its MD, if the orbit of galaxy~$j$ around its MD is Keplerian. 
Thus, $\Theta_1 \ge 0$ implies that galaxy~$j$ is most likely bound to its MD, while $\Theta_1 < 0$ means that galaxy~$j$ is unbound, as also detailed in \cite{bib:Karachentsev1999} and \cite{bib:Karachentsev2013}.

\begin{figure*}
    \centering
    \includegraphics[width=0.32\linewidth]{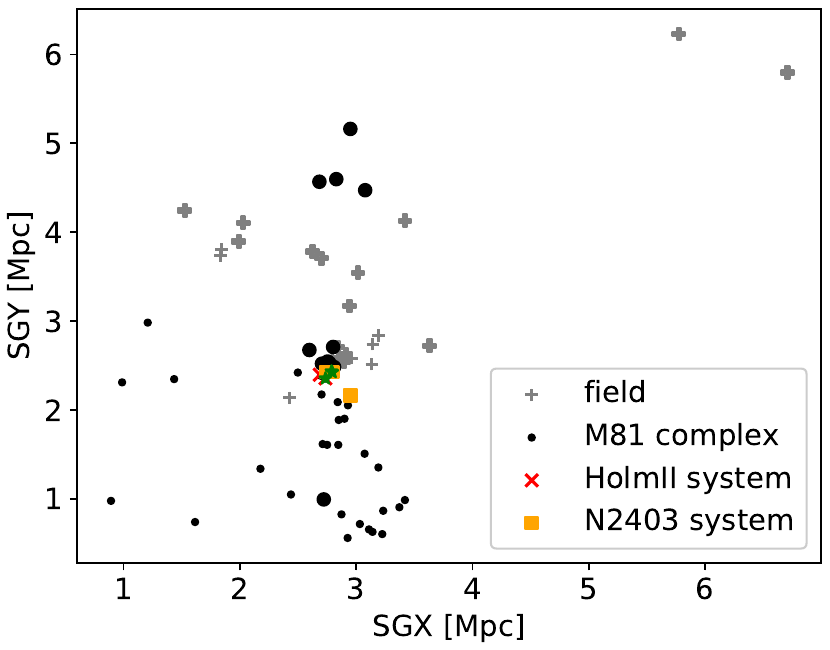} \hfill 
    \includegraphics[width=0.335\linewidth]{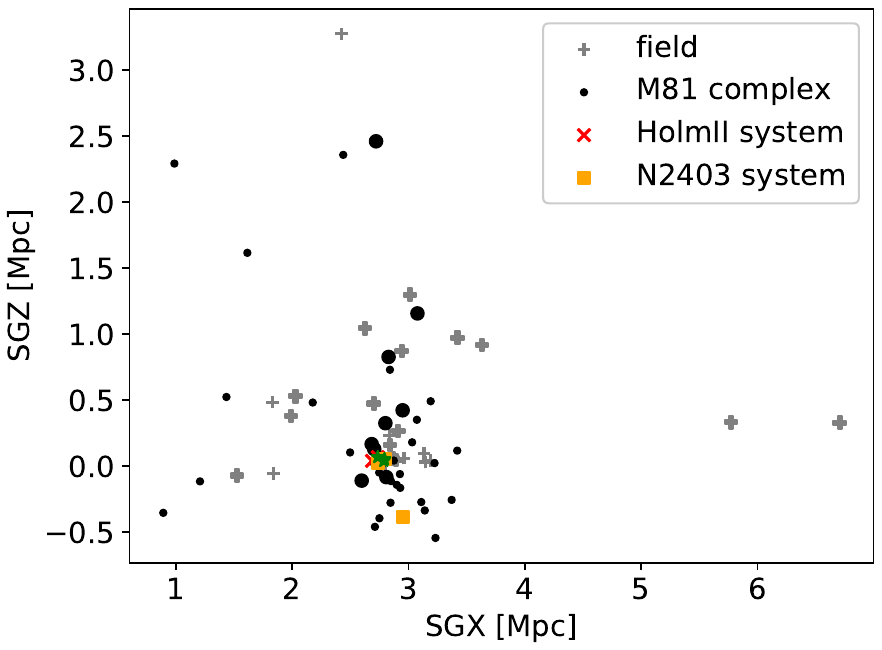} \hfill
    \includegraphics[width=0.335\linewidth]{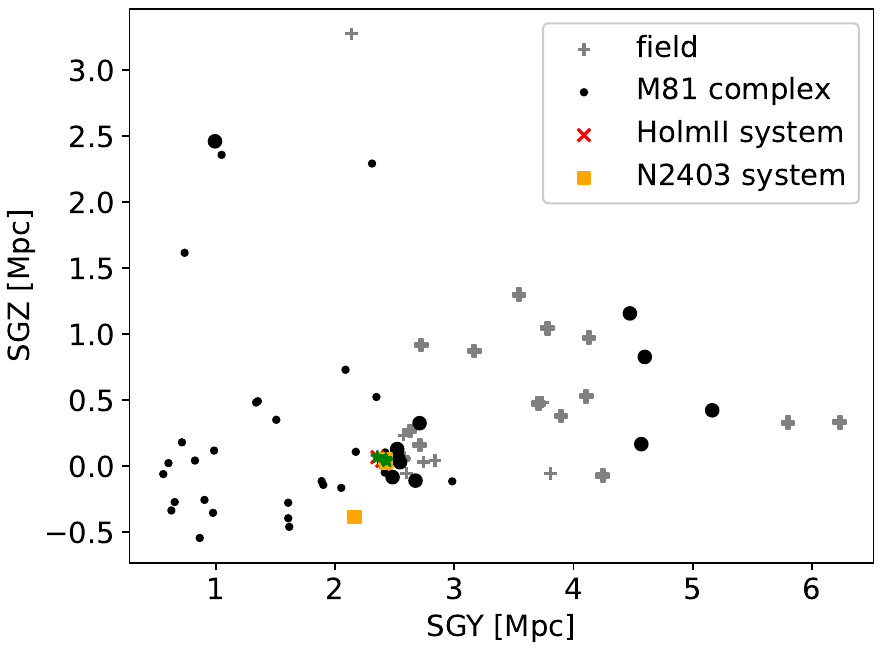}
    \caption{\it{Locations of the 72 galaxies having velocity measurements: physical positions in supergalactic coordinates in the SGX-SGY-plane (left), SGX-SGZ-plane (centre), and SGY-SGZ-plane (right). Galaxies in the environment are marked as ``field" (grey pluses), galaxies having M~81, M~82, NGC~3077 or NGC~2976 as their major disturber are summarised under ``M81 complex" (black dots), the galaxy attributed to Holm~II as its major disturber and Holm~II itself are called the ``HolmII system" (red xs), NGC~2403 and two satellites galaxies are called the ``N2403 system" (yellow squares). M~81, Holm~II, and NGC~2403 are additionally highlighted by green stars. Larger symbols indicate galaxies having $\Theta_1 < 0$ and thus being unbound.}}
    \label{fig:data_plot}
\end{figure*}

Grouping the 58 galaxies can now be performed by noting that the six brightest galaxies in the $B$-band are M~81, NGC~2403, M~82, NGC~3077, NGC~2976, and Holm~II, with M~81, M~82, NGC~3077, and NGC~2976 being the four central galaxies of the M~81 complex. 
Uniting those galaxies having one of these four central galaxies as their MD yields 31 galaxies attributed to the M~81 complex. 
Only one galaxy, KDG~052, is attributed to Holm~II as MD and two galaxies form a system around NGC~2403, DDO~44 and NGC~2366, having NGC~2403 as their MD. 
Out of the 31 galaxies in the M~81 complex, 5 galaxies, all with M~81 as their MD, have $\Theta_1 < 0$\footnote{There are six more galaxies with other distance measures having $\Theta_1 < 0$, so it may be worth obtaining TRGB-based distances for those.}. 
Apart from them, there are 14 additional galaxies with other MDs beyond a distance of 1.5~Mpc from M~81 which also have $\Theta_1 < 0$. For each galaxy in our data set, its $B$-band magnitude was measured as well. We summarise all information available for the 72 galaxies in Tab.~\ref{tab:M81-group} in an appendix. 
To visualise the data set, we choose supergalactic coordinates, see Fig.~\ref{fig:data_plot}.

\subsection{Comparison to our previous data set}
\label{sec:comparision_Mueller}

All 21 galaxies of our previous data set discussed in \cite{bib:Wagner2025}, originally assembled by \cite{bib:Mueller2024}, are contained in the new data set, d1012+64 being called UGC05497. 
Table~\ref{tab:M81-group} indicates the membership of a galaxy in the \cite{bib:Mueller2024} data set with a star-superscript behind the name of the galaxy.
Hence, the 21 galaxies are all at distances much less than 1~Mpc from M~81. 

For KDG~063 and HS~117, no velocities were given in the new data set. 
KDG~063 is missing a far-UV flux, so that the previous velocity measurement may be a spurious detection of local Galactic emission. 
HS~117 shows a far-UV flux, yet only a faint one, requiring a follow-up observation.

The eight satellites in the \cite{bib:Mueller2024} data set beyond the second-turnaround radius of 230~kpc all have $\Theta_1$-values larger or equal than 1.0, so we consider them to be bound to the M~81 complex for now.

\section{Embedding into the local environment}
\label{sec:embedding_environment}

To understand the structure of the Hubble flow around the M~81 complex, we analyse its embedding into the local environment. We collected 103 galaxies with TRGB- and SBF-based distances with a distance of less than 8~Mpc from us and 5.12~Mpc around M~81, a supergalactic latitude smaller than $10^\circ$, and a supergalactic longitude between $0^\circ$ and $80^\circ$ for this task. 
This sample, restricted to SGL~$< 65^\circ$, is highly complete down to $\log(L_K) = 7.0$.
The larger surroundings also point at the mini void behind the M~81 complex as seen from us.
Table~\ref{tab:M81-environment} lists all galaxies used in this section.

In \cite{bib:Mueller2024}, the flattened structure of the satellite galaxies within a radius of about 250~kpc around the barycentre is studied using the galaxies marked with a star in Table~\ref{tab:M81-group}.
Additional galaxies (not listed in Table~\ref{tab:M81-group} because they still lack velocity measurements) are also used to constrain the spatial structure of the M~81 complex in \cite{bib:Mueller2024}. 
The direction of the smallest extent of the M~81 complex in supergalactic coordinates was found to be $\boldsymbol{n} = \left(-0.1314, +0.0374, +0.9906 \right)$. 
Excluding four potential outliers, one of them being HS~117, the resulting vector aligns even more with the SGZ-axis.

In comparison, we perform a singular value decomposition (SVD) to determine the axes of largest and smallest extent of the larger volume covered by our dataset. 
To investigate the robustness of the result, we first anchor the SVD around the mean location, then around the median location of all galaxies in Table~\ref{tab:M81-environment}, and, at last, around the centre of mass of the six galaxies with highest $L_K$, IC~342, NGC~2403, M~81, NGC~2787, NGC~4258, and NGC~4736. 
We obtain the direction of smallest extent for the mean of all galaxies as $\boldsymbol{n}_\mathrm{mean} = \left(0.0498 , -0.1439, 0.9883\right)$~Mpc, for the median $\boldsymbol{n}_\mathrm{median} = \left(0.0500, -0.1441, 0.9883\right)$~Mpc, and for the centre of mass of the six most luminous galaxies $\boldsymbol{n}_\mathrm{com} = \left(0.0368, -0.1567, 0.9870\right)$~Mpc. 
Thus, the alignment of these directions with respect to the SGZ-axis is recovered in the larger volume as well.

Moreover, we also analyse the dependence on the volume covered by the galaxies and the selection of galaxies by restricting the dataset to the 89 galaxies within 3~Mpc from M~81 and to the 34 galaxies within 0.25~Mpc around M~81, which is approximately the volume covered by \cite{bib:Mueller2024}, yet sampled with slightly different galaxies.
While the former yields a good alignment to the previous three directions of smallest extent, $\boldsymbol{n}_\mathrm{<3Mpc} = \left(-0.0392, -0.1887, 0.9812\right)$~Mpc, the inner core is not well aligned anymore with its $\boldsymbol{n}_\mathrm{<0.25Mpc} = \left(-0.4602,  0.2073,  0.8633\right)$~Mpc.

\begin{figure*}
    \centering
    \includegraphics[width=0.5\linewidth]{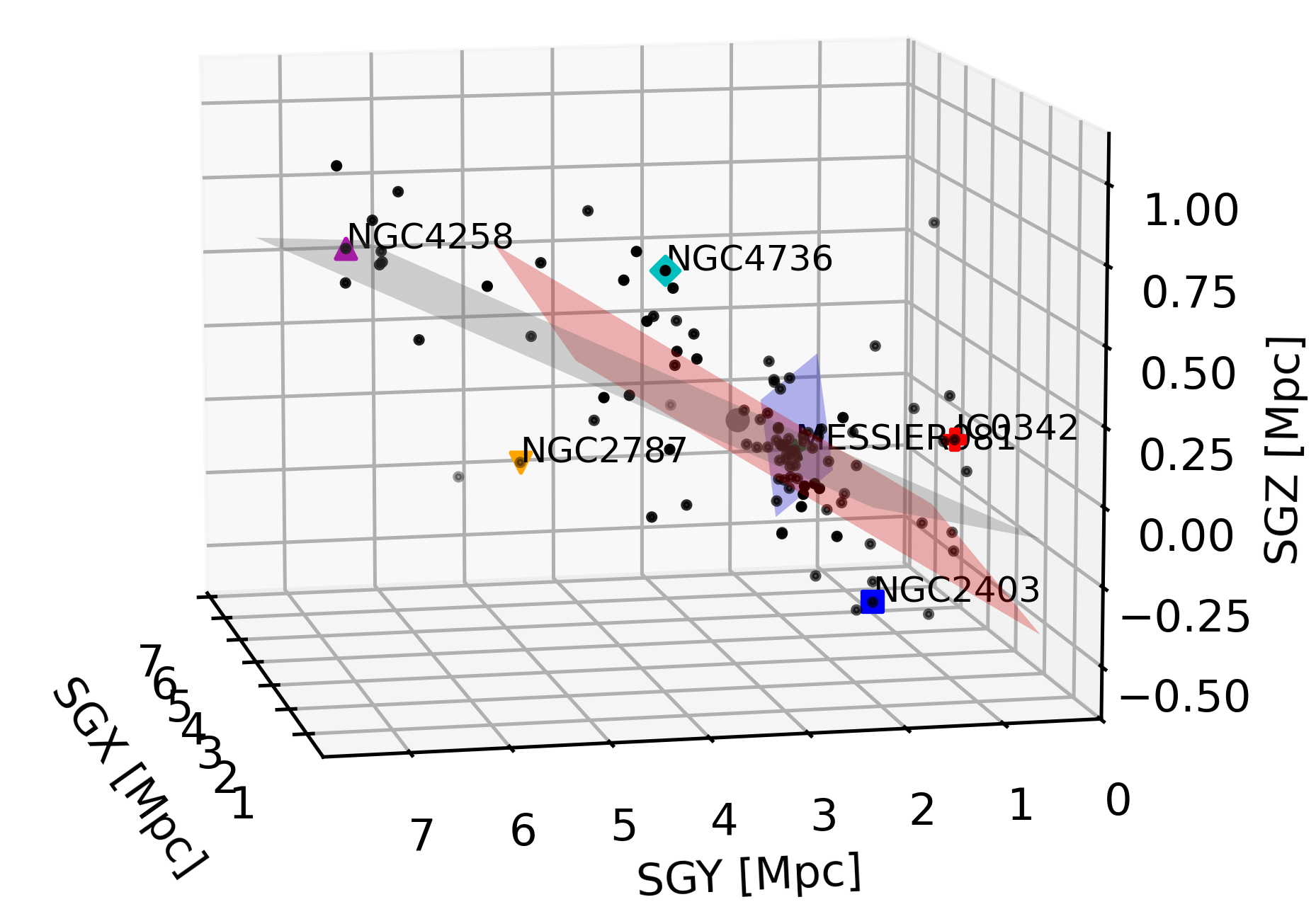} \hfill
    \includegraphics[width=0.43\linewidth]{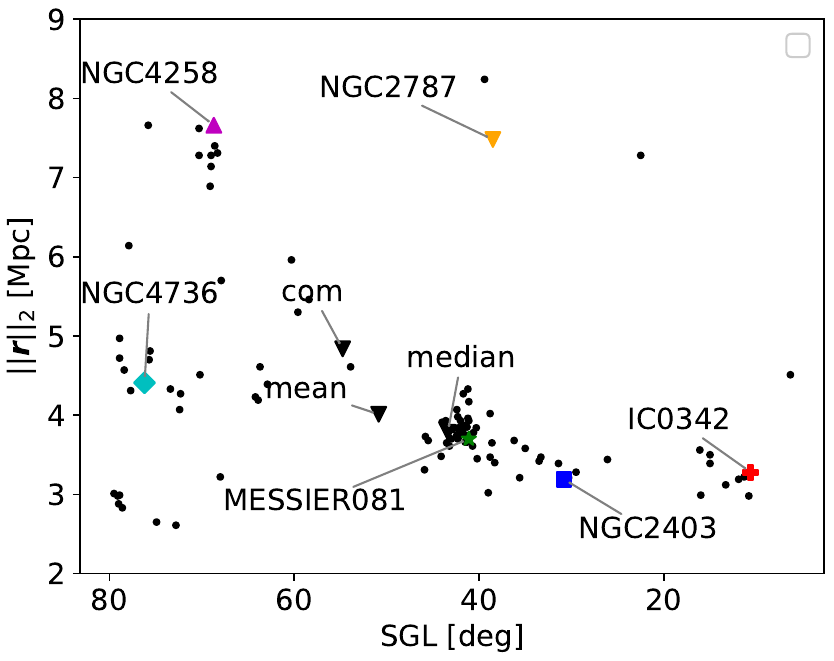}
    \caption{\it{(Movie online, left plot) Filamentary structure of the local M~81-group including all galaxies of Table~\ref{tab:M81-environment}: Determining the directions of smallest extent for all 103 galaxies anchored at their mean location, the normal vector to the grey plane is obtained. An analogous calculation for the volume in a 3~Mpc radius around M~81 yields the red plane and its normal direction. For the volume in the 0.25~Mpc radius around M~81, the blue plane and its normal direction are obtained. The former two planes are becoming more aligned to the SGX-SGY-plane for increasing volume size. For further orientation, the most luminous galaxies in our data set are marked in colour and their names are attached. The right plot shows a projection into the plane of SGL and the distance from us, also showing the locations of the mean, median and the centre of mass of the six most luminous galaxies.}}
    \label{fig:environment_embedding}
\end{figure*}

Fig.~\ref{fig:environment_embedding} shows the resulting planes of maximum extent for the entire data volume, the galaxies in a 3~Mpc radius around M~81, and the galaxies in a 0.25~Mpc radius around M~81. 
As can be read off the plot, and more clearly from the rotating plot in the online movie, the planes become more aligned with increasing volume size.
The selection of galaxies by \cite{bib:Mueller2024} is already sufficient to arrive at the larger-scale alignment roughly oriented parallel the SGX-SGY plane in good agreement with many other works detecting sheet-like structures in the Local Universe in similar directions, for instance \cite{bib:McCall2014}, \cite{bib:Anand2019}, or \cite{bib:Peebles2023}. 

Beyond that, we observe a filamentary structure around the M~81 complex on a smaller scale.
This structure is not captured by a single plane, but rather by two filament arms, one from M~81 towards NGC~4258 and another to NGC~4736. 
The grey plane of maximum extent contains NGC~4258 (also known as M~106, with its water megamaser that has been used extensively to calibrate the cosmic distance ladder, see, for instance, \cite{bib:Tully2023} for a recent discussion), yet it does not contain NGC~4736 (also known as Messier~94) and its satellites. 
Restricting the volume to a radius of 3~Mpc around M~81 and thereby excluding NGC~4258 and its satellites from the analysis, the red plane is lifted slightly towards NGC~4736, however, still not capturing this filamentary part of M~81's environment. 

Connecting the local environment of the M~81 complex to the filaments found by \cite{bib:Raj2024}, the M~81 complex lies in their ``filament~10", which consists of 63 galaxies with M~81 (called NGC~3031) marked as the major galaxy in the centre of this filament and Messier~101 (called NGC~5457) being at the right end. 
The next galaxy close to M~81 is NGC~672 attributed to ``filament 0" at the crossing of various filaments.
The larger filamentary structure in \cite{bib:Raj2024} seems to share its direction with our grey plane, but occurs on a larger scale.
Vice versa, the filamentary structures around the M~81 group captured in our work is yet on a slightly larger scale than the M~81 satellite alignments found by \cite{bib:Mueller2024}. 
On the whole, irrespective of the scale we look at beyond a radius of 0.25~Mpc around M~81, the filaments seem to have a preferred alignment direction approximately perpendicular to the SGX-SGY plane.

\section{Galaxy infall models}
\label{sec:galaxy_infall}

\subsection{Velocity corrections}
\label{sec:velocity_corrections}

Before we apply the major and minor infall models, as set up in \cite{bib:Karachentsev2006}, to the velocities of the galaxies, possible corrections for relative mutual motions are in order. 
The impact of the observers' own motion on the infall models is still an open question to be investigated in this work because the measured velocities towards galaxies around M~81 contain a contribution of the solar motion with respect to the Milky Way centre and the Milky Way's motion towards the Local Group centre. 
Assuming that both, the M~81 complex and the Local Group both move towards the Virgo cluster, correcting the measured velocities for the solar motion within the LG is sufficient to account for the relative motion between us and the M~81 complex. 

Thus, we can either use the heliocentric velocities, $v_\mathrm{hel}$, or take into account the solar motion in the Local Group's gravitational potential. 
The latter are called $v_\mathrm{LG}$ in the following. We convert the measured $v_\mathrm{hel}$ into $v_\mathrm{LG}$ via
\begin{align}
v_\mathrm{LG} = v_\mathrm{hel} &+ v_\odot \big( \sin (b) \sin(b_\odot) + \cos(b) \cos(b_\odot) \cos(l-l_\odot) \big) \;,  
\label{eq:v_LG}
\end{align}
in which $b$ and $l$ denote the latitute and longitude of the galaxy in the Galactic coordinate system (following the IAU’s 1958 definition).
For consistency with the data base of \cite{bib:Karachentsev2013}, we also use the solar apex with respect to the Local Group rest frame
\begin{equation}
l_\odot = \left(93 \pm 2 \right)^\circ\,, \;\; b_\odot = \left(-4 \pm 2 \right)^\circ\,, \;\; v_\odot = \left(316 \pm 5 \right)~\mathrm{km}/\mathrm{s}
\label{eq:apex}
\end{equation}
as determined in \cite{bib:Karachentsev1996}. 
Although this measurement goes back to the last century, it is still used as reference for the NASA Extragalactic Database\footnote{\url{https://ned.ipac.caltech.edu/help/velc_help.html}} and has been re-measured over the years, for instance, in \cite{bib:Tully2008} or \cite{bib:Makarov2025}, showing a high degree of consistency.
When converting $v_\mathrm{hel}$ to $v_\mathrm{LG}$, the uncertainties in the solar apex need to be taken into account as well, resulting in larger error bars for the LG-based velocities, as detailed in the following sections.

\subsection{Barycentre of the M~81 complex}
\label{sec:barycentre}

Since the M~81 complex is a group of galaxies, we need to determine its barycentre, $\boldsymbol{r}_\mathrm{c}$, which serves as the centre for all infalling galaxies and the Hubble flow. 
In \cite{bib:Wagner2025}, we assumed M~81 to be the barycentre as a first ansatz and considered galaxies within a radial distance of 230~kpc, the extent of the second turnaround radius, to be gravitationally bound. 
Assuming uncertainties in the distance measures to be 5\% for TRGB-based distances and 30\% for all other methods, we find that this ansatz is still a good approximation to the true barycentre because the next three central, brightest galaxies, M~82, NGC~3077, and NGC~2976, all have three-dimensional distances to M~81 which are compatible with zero within their observational uncertainties.

Nevertheless, we investigate a possible refinement of the barycentre position.
In \cite{bib:Karachentsev2006} two different methods were used to determine the barycentre: 
(1) applying the infall models to all galaxies with respect to a variable $\boldsymbol{r}_\mathrm{c}$ and determining $\boldsymbol{r}_\mathrm{c}$ as the location which minimises the velocity dispersion around a Hubble flow fit (see Sect.~\ref{sec:impact_parameter}); 
(2) assuming that the mass of the galaxies is proportional to their $K$-band luminosity, $L_K$, with the same mass-to-light ratio, $\Upsilon \equiv m_j/L_{Kj}$, for all galaxies~$j$.
 {This is motivated in \cite{bib:Makarov2011} and \cite{bib:Tully2015} from the fact that the observed infrared flux in the local cosmic volume is hardly influenced by dust or young stellar contents, so the same mass-to-light ratio can be used to estimate the stellar mass for all galaxies irrespective of their type and morphology. Assuming that the dark-matter content scales in the same manner,}  
the three-dimensional centre of mass is calculated from the observed angles on the sky and distance
\begin{equation}
\boldsymbol{r}_\mathrm{c} = \sum \limits_{j=1}^{n_\mathrm{g}} \frac{\Upsilon L_{Kj}}{\Upsilon L_{K \mathrm{tot}}} \boldsymbol{r}_j \;, \quad L_{K \mathrm{tot}} = \sum \limits_{j=1}^{n_\mathrm{g}} L_{K j} \;,
\label{eq:rc}
\end{equation}
in which $n_\mathrm{g}$ denotes the number of all galaxies gravitationally bound to the M~81 complex. 
Since $\Upsilon$ cancels out, we do not require a specific value for the mass-to-light ratio to be known, only the weaker assumption that $\Upsilon$ is the same for all galaxies involved. 
As noted in \cite{bib:Wagner2025}, this formula requires the group's mass to be contained in the galaxies without additional, group-scale dark-matter halos. 

Similar to the LG, \cite{bib:Karachentsev2006} simplified Eq.~\eqref{eq:rc} to only include the two brightest galaxies in the group, M~81 and M~82, such that $\boldsymbol{r}_\mathrm{c}$ is located on their connection line. 
We repeat their calculation with the updated observations and also extend the study to the barycentre determined from the four brightest central galaxies in the M~81 complex. 
We retrieve their $K$-band luminosities from \cite{bib:Karachentsev2013} as $\log(L_K(\mathrm{M~81})) = 10.95$, $\log(L_K(\mathrm{M~82})) = 10.59$, $\log(L_K(\mathrm{N~3077})) = 9.57$, and $\log(L_K(\mathrm{N~2976})) = 9.44$.
As the central galaxies of the M~81 complex fulfil the small angle approximation, we calculate the line-of-sight velocity component of the barycentre by replacing $\boldsymbol{r}$ by $\boldsymbol{v}_\mathrm{hel}$ in Eq.~\eqref{eq:rc}.
Table~\ref{tab:barycentre} summarises all results. 

\begin{table*}
\centering
\begin{threeparttable}
\caption{ \centering \it{Barycentres of the M~81 complex assuming M~81 (first row) and the next three brightest central galaxies (accumulating one galaxy after the other from the second to the fourth row) constrain its position ($\alpha_\mathrm{c}$, $\delta_\mathrm{c}$, and the distance $r_\mathrm{c}$ of the barycentre to us as observers) and its line-of-sight velocity component ($v_\mathrm{hel c}$), uncertainties in the distance ($\Delta r_\mathrm{c}$) and line-of-sight velocity ($\Delta v_\mathrm{hel c}$) are given by linear propagation of independent uncertainties. The latter only contain measurement imprecisions and no systematic biases due to potential offsets between dark and luminous matter. The last three columns show the three-dimensional distance between $\boldsymbol{r}_\mathrm{c}$ and $\boldsymbol{r}_\mathrm{M~81}$, the line-of-sight velocity corrected for the sun's motion in the LG, $v_\mathrm{LG}$, and its uncertainty $\Delta v_\mathrm{LG}$.  {The last row lists the barycentre properties for a varying mass-to-light ratio, see text.}}}
\label{tab:barycentre}
\begin{tabular}{lccccccccc}
\hline
Galaxies & $\alpha_\mathrm{c}$ & $\delta_\mathrm{c}$ & $r_\mathrm{c}$ & $\Delta r_\mathrm{c}^*$ & $v_\mathrm{hel c}$ & $\Delta v_\mathrm{hel c}^\dagger$ & $d(M~81)$ & $v_\mathrm{LG c}$ & $\Delta v_\mathrm{LG c}^\dagger$ \\
 & $\left[ \mathrm{deg} \right]$ & $\left[ \mathrm{deg} \right]$ & $\left[\mathrm{Mpc}\right]$ & $\left[\mathrm{Mpc}\right]$ & $\left[\mathrm{km}/\mathrm{s}\right]$ & $\left[\mathrm{km}/\mathrm{s}\right]$ & $\left[\mathrm{Mpc}\right]$ & $\left[\mathrm{km}/\mathrm{s}\right]$ & $\left[\mathrm{km}/\mathrm{s}\right]$\\
\hline 
M~81 & 148.889583 & 69.066667 & 3.70 & 0.19 & -38 & 1 & 0.00 & 104 & 10\\
... \& M~82 & 148.915412 & 69.253801 & 3.67 & 0.14 & 29 & 1 & 0.03 & 172 & 10\\
... \& N~3077 & 148.969617 & 69.239139 & 3.68 & 0.14 & 29 & 1 & 0.03 & 171 & 10\\
... \& N~2976 & 148.925496 & 69.211995 & 3.68 & 0.13 & 28 & 1 & 0.02 & 171 & 10\\
\hline
 {M~81 \& M~82} &  {148.897594} &  {69.124703} &  {3.69} &  {0.17} &  {-17} &  {5} &  {0.01} &  {125} &  {11} \\
\hline
\end{tabular}
\begin{tablenotes}
\item $^*$ determined from distance uncertainties only, assuming 10\% or 20\% uncertainty in the $L_K$, changes in $\Delta r_\mathrm{c}$ remain below the precision given here.  {Uncertainties in the mass-to-light ratio in the last row are taken into account as given by \cite{bib:Karachentsev2021}.}
\item $^\dagger$ determined from distance uncertainties only, assuming 10\% uncertainty in the $L_K$: $\Delta v_\mathrm{hel c} = \left[1,7,7,6 \right]$~km/s, for 20\%, we obtain: $\Delta v_\mathrm{hel c} = \left[1,13,13,13 \right]$~km/s, analogously for 10\% uncertainty in $L_K$: $\Delta v_\mathrm{LG c} = \left[10,12,12,12 \right]$~km/s, for 20\%: $\Delta v_\mathrm{LG c} = \left[10,17,16,16 \right]$~km/s.  {Including the uncertainties in the mass-to-light ratio in the last row, assuming additional 10\% uncertainty in the $L_K$: $\Delta v_\mathrm{hel c} = 6$~km/s and $\Delta v_\mathrm{LG c} = 11$~km/s, and for 20\%: $\Delta v_\mathrm{hel c} = 7$~km/s and $\Delta v_\mathrm{LG c} = 12$~km/s.}
\end{tablenotes}
\end{threeparttable}
\end{table*}

The uncertainties in $r_\mathrm{c}$, the absolute value of the barycentre distance to us, and $v_\mathrm{LGc}$, the absolute value of the line-of-sight velocity component corrected for our motion within the LG, were obtained by a linear propagation of uncertainty, taking into account the uncertainties in the line-of-sight distances, line-of-sight velocities, and the uncertainties in the apex parameters when correcting for our motion in the LG.
The detailed equations for the propagation of all uncertainties are given in Appendix~\ref{app:error_propagation}.

 {At last, we investigate a varying dark-matter content for the spiral M~81, de Vaucouleurs morphological type 1 or 2, and the spiral but edge-on starburst M~82, de Vaucouleurs morphological type 5, 6 or 7, according to \cite{bib:Karachentsev2021}. Then, $\Upsilon(\text{M81}) = (73\pm15)~M_\odot/L_\odot$ and $\Upsilon(\text{M82}) = (17.4\pm2.8)~M_\odot/L_\odot$. Calculating the barycentre properties including the varying $\Upsilon$ is summarised in the last row of Table~\ref{tab:barycentre}. Only the amplitudes and directions of the velocity vectors in the heliocentric and the Local-Group frame change significantly, while the position of the centre still coincides with the one of M~81 within uncertainty bounds.}

We thus conclude that the position of the centre of mass remains robustly located at the position of M~81 within the uncertainty bounds.
Yet, $v_\mathrm{hel}$ of the barycentre changes beyond these limits, which will become important for the infall models in Sect.~\ref{sec:infall_models}. 
Only taking into account M~82 as the second-brightest central galaxy is sufficient to constrain the properties of the barycentre under the prerequisites stated above because the properties, including $v_\mathrm{hel}$, remain stable when including fainter central galaxies.
 {Allowing for a varying mass-to-light ratio to account for different dark-matter contents of M~81 and M~82 according to heuristics found in \cite{bib:Karachentsev2021}, the velocity vectors tend to align more towards the ones of M~81. Yet, this barycentre does not minimise the velocity dispersion around the Hubble flow like the one with constant $\Upsilon$ does, as shown in a systematic, dark-matter-model-independent analysis in \cite{bib:Karachentsev2006}. Moreover, determining the barycentre based on a minimisation of the velocity dispersion of the bound galaxies, similar to \cite{bib:Karachentsev2009}, we recover the same barycentre within uncertainty bounds as the one listed in the second row of Table~\ref{tab:barycentre} and the one of \cite{bib:Karachentsev2006}. Due to this overall consistency, we thus choose the barycentre as obtained in the second line of Table~\ref{tab:barycentre} in the following.}


\subsection{Infall models}
\label{sec:infall_models}

After defining a centre with respect to which the infall models are calculated and setting up the velocity corrections due to potential biases, we now introduce the infall models as already detailed in \citep{bib:Karachentsev2006} and \cite{bib:Wagner2025}. 

\begin{figure}[t!]
    \centering
    \includegraphics[width=0.75\linewidth]{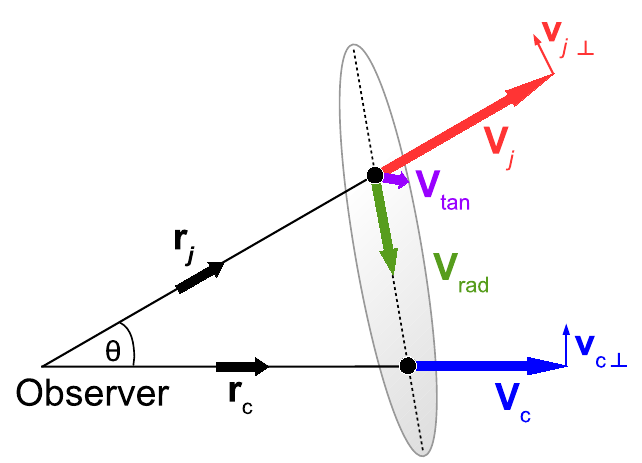}
    \begin{tabular}{cc}
\hline
      Model & Components set to zero \\
    \hline
   Minor  &  $v_{1,\perp}$ and $v_{2,\perp}$   \\
    Major  &  $v_{c,\perp}$ and $v_{\text{tan}}$       
    \\
\hline
    \end{tabular}
    
\caption{ {Relative location and motion of a galaxy~$j$ at distance $r_j$ from an observer with respect to the centre of mass at distance $r_\mathrm{c}$. The vector $\boldsymbol{v}_j$ represents the line-of-sight velocity of galaxy~$j$, while $\boldsymbol{v}_\mathrm{c}$ is the line-of-sight velocity of the barycentre. The angle $\theta$ is measured between $\boldsymbol{r}_j$ and $\boldsymbol{r}_\mathrm{c}$. The prerequisites for the two infall models to accurately describe the radial velocity $\boldsymbol{v}_\mathrm{rad}$ are listed in the table.}}
    \label{fig:rel_motion}
\end{figure}

In the following, all infall models approximate the radial velocity component between the centre of mass of the M~81 complex and the surrounding satellite galaxies. 
To simplify the notation compared to \cite{bib:Wagner2025}, we denote the absolute values of the line-of-sight distance and velocity components as $r$ and $v$, respectively. 
Subscripts $c$ and $j$ denote the group's barycentre and a galaxy~$j$, respectively, such that $r_{\mathrm{c}j}$ denotes the absolute value of the three-dimensional distance between the barycentre and galaxy~$j$.
The angular distance on the sky between the barycentre and a galaxy~$j$ is denoted by $\theta_{\mathrm{c}j}$.
 {Fig.~\ref{fig:rel_motion} summarises the notation and directions of vectors.} 

The \textbf{minor infall model} approximates the amplitude\footnote{Our notation is the same as in \cite{bib:Wagner2025}, with $\left| \cdot \right|$ denoting the signed amplitude of the radial velocity vector with fixed direction towards the barycentre, positive sign in this direction and negative sign otherwise. The absolute value of a vector is defined as $\Vert \cdot \Vert_2$.} of the radial infall velocity of galaxy~$j$ onto the barycentre as:  
\begin{align}  
\left| \boldsymbol{v}_{\mathrm{min}j} \right| = \frac{v_{\mathrm{c}} {r}_{\mathrm{c}} + v_{j}  {r}_{j} - \cos \theta_{\mathrm{c}j}  ( v_j  {r}_{\mathrm{c}} + v_\mathrm{c} {r}_{j} )}{r_{\mathrm{c}j}} \;.  
\label{eq:v_min}  
\end{align}  

This approximation treats the barycentre and galaxy~$j$ symmetrically and is an accurate representation of the radial infall velocity if the velocity components perpendicular to the line of sight are negligible.

The \textbf{major infall model} is an asymmetric approximation to the radial infall velocity of galaxy~$j$ onto the barycentre:
\begin{align}  
\left| \boldsymbol{v}_{\mathrm{maj}j} \right| = \frac{v_j - v_\mathrm{c} \cos \theta_{\mathrm{c}j}}{r_j - r_\mathrm{c} \cos \theta_{\mathrm{c}j}} r_{\mathrm{c}j} \;.  
\label{eq:v_maj}  
\end{align}  
It can be derived from projecting the radial velocity onto the infalling galaxy's line of sight. 
This model becomes an accurate representation of the radial infall velocity for a gravitational bound structure only moving with the cosmic expansion as a whole and containing only radially infalling galaxies.
 {The table included in Fig.~\ref{fig:rel_motion} lists the accuracy conditions once again.}

The angular separation of the galaxies on the sky to the barycentre of the extended M~81 complex is about $44^\circ$, such that the small-angle approximation for both infall models by simply subtracting the line-of-sight velocity components of the barycentre and galaxy~$j$ is invalid. 
The latter is only applicable to the galaxies within the second turnaround radius which span a maximum angle of less than $2.3^\circ$ on the sky, as detailed in \cite{bib:Wagner2025}. 

As shown in \cite{bib:Wagner2025} and \cite{bib:Benisty2025} by means of simulations, the minor infall statistically underestimates the radial velocity dispersion, while the major infall is an overestimate of the radial velocity dispersion for most of the groups tested. 
Applying both infall models to a cosmic structure, lower and upper bounds on the radial velocity dispersion are obtained. 

Due to the lack of precise proper motions, only the line-of-sight velocity component is employed, such that the infall models are biased, as detailed in \cite{bib:Wagner2025}.  
Moreover, while three-dimensional distance observations are possible nowadays, the celestial angles on the sky can be measured to much higher precision than the distance along the line of sight. 
Hence, the major sources of uncertainties in the resulting approximations to the radial infall velocity are the line-of-sight distances. 

Fig.~\ref{fig:individual_infall} in the appendix shows all galaxy infall velocities for the minor (blue) and major infall models (red) and the simple difference of the line-of-sight velocities as the leading-order approximation to the radial velocity between each galaxy~$j$ and M81, called $v_\mathrm{app}$ in \cite{bib:Wagner2025} (black). 
The galaxies are sorted according to their three-dimensional distance to M~81. The closest is on the left, the farthest is on the right. 
The major infall model yields velocity amplitudes larger than 1000~km/s for 13 of the galaxies, which is why these values are not shown in the plot. 
While this figure is produced using the barycentre on the connection line between M~81 and M~82 and taking into account the observers' motion within the Local Group, we systematically investigate the impact of each of these assumptions on the infall models and on the Hubble flow fit in the next section.

\section{Hubble flow fit}
\label{sec:Hubble_flow}

\subsection{Fitting function and uncertainty propagation}
\label{sec:fitting_function}

With the infall velocities set up in Sect.~\ref{sec:galaxy_infall} and the independently determined distances for all galaxies to the barycentre of the M~81 complex, denoted as $\Vert \boldsymbol{r}_{\mathrm{c}j} \Vert_2 \equiv r_{\mathrm{c}j}$ for each galaxy~$j$, we plot the Hubble diagram for the M~81 group.
For the Hubble flow fit to the Hubble diagram, we use the approximation from \cite{bib:Peirani2006,bib:Peirani2008,bib:Penarrubia2014,bib:Benisty2025b}:
\begin{equation}
\left| \boldsymbol{v}_{\mathrm{inf}j} \right| = \left| \boldsymbol{v}_{\mathrm{inf}j} \right|\left(r_{\mathrm{c}j}\right) = \kappa H_0 r_{\mathrm{c}j} - 1.1 \sqrt{\frac{G M}{r_{\mathrm{c}j}}} \;,
\label{eq:HFF}
\end{equation}
in which $\kappa$ is a scaling factor in front of the Hubble constant $H_0$, $G$ is the gravitational constant, and $M$ is the total mass of the M~81 complex. 
In \citet{bib:Penarrubia2014}, $\kappa\approx1.4$, but varying values can be found in the literature, depending on the embedding background~\citep{bib:DelPopolo2021,bib:DelPopolo2022}. 
The fit thus yields $H_0$ and $M$ from the infall velocities and the observed distances of all galaxies in the M~81 complex. 
The turnaround, $|\boldsymbol{v}_{\mathrm{inf}j}| = 0$, from Eq.~\eqref{eq:HFF} yields the known enclosed mass relation from~\citet{bib:Sandage1986}:
\begin{equation}
M = 1.8 \cdot 10^{12} \left(\frac{H_0}{70\, \text{km/s/Mpc}}\right)^2 \left( \frac{r_{\text{ta}}}{1\, \text{Mpc}}\right)^3
\label{eq:hubble_flow_enclosed_mass},
\end{equation}
where $r_{\text{ta}}$ is the turnaround distance of the structure.

Before we determine the best-fit $H_0$ and $M$, we investigate the impact of shifting the barycentre from M~81 to the point on the connection line between M~81 and M~82 determined in Sect.~\ref{sec:barycentre}, as well as the impact that the corrections to the velocities for our motion in the Local Group detailed in Sect.~\ref{sec:velocity_corrections} have. 
Since the Hubble flow also starts at a specified minimum distance from the barycentre, $d_\mathrm{min}$ and extends to a maximum distance from the barycentre, $d_\mathrm{max}$, the impact of these fitting parameters also needs to be investigated.  
For all analyses, we only employ the 58 galaxies having TRGB-based distances with the highest measurement precision out of all distance-measure methods. 

In order to obtain the most realistic estimates of uncertainties, for each galaxy~$j$ having a TRGB-based distance, we draw 10,000 samples from a Gaussian distribution around the observed $r_j$ and $v_{\mathrm{hel}j}$, respectively, using $\Delta r_j = 0.05\, r_j$ and $\Delta v_{\mathrm{hel}j}$ as widths. 
Subsequently, for each of the 10,000 sample data sets à 58 galaxies, we determine the barycentre (either being the position of M~81 in this sample data set, or being the point on the connection line between the M~81- and M~82-sample detailed in Sect.~\ref{sec:barycentre}).
We then calculate the infall velocities of all 58 galaxies towards this barycentre for each of the 10,000 data sets. 
In this way, the standard deviation of the 10,000 samples around the mean distance of galaxy~$j$ to the barycentre serves as the confidence bound on the TRGB-based distances. 
For the minor infall velocity, we analogously set up its confidence bound based on the standard deviation around the mean minor infall velocity for each galaxy. 
For the major infall velocity, we need to set up its confidence bound based on the 0.159 and 0.841 quantiles, which corresponds to the 1-$\sigma$ bounds around the mean. 
This is the case due to chance alignments between galaxies and the barycentre that cause the denominator in Eq.~\eqref{eq:v_maj} to diverge. 

\begin{figure*}
\centering
\includegraphics[width=0.4\linewidth]{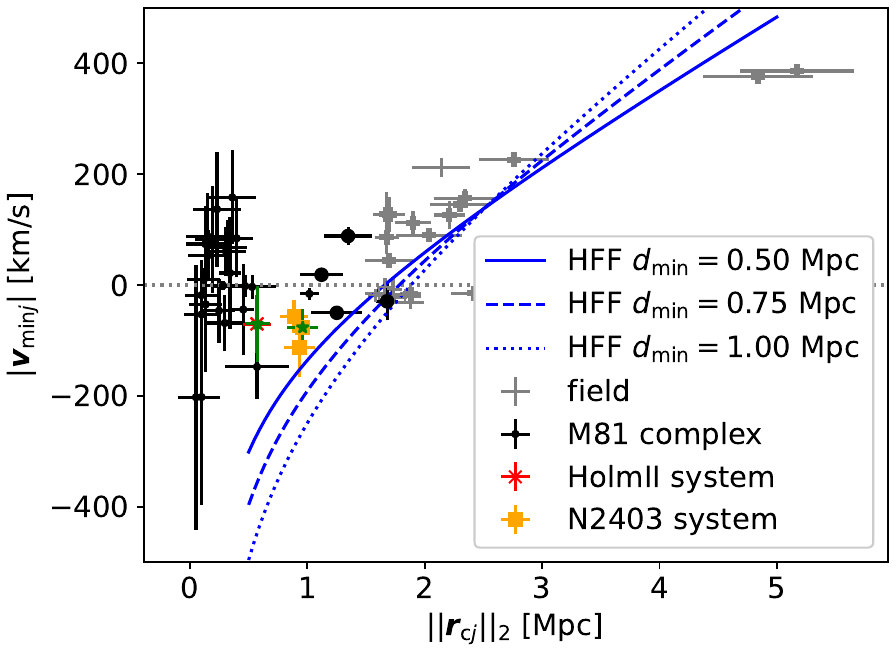} \hspace{3ex}
\includegraphics[width=0.4\linewidth]{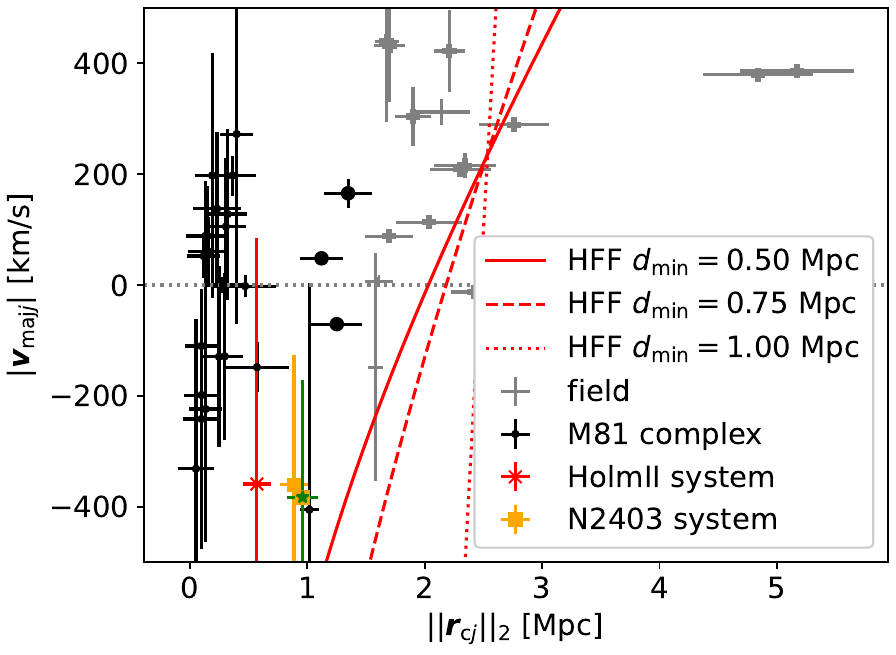} \\[1ex]
\includegraphics[width=0.4\linewidth]{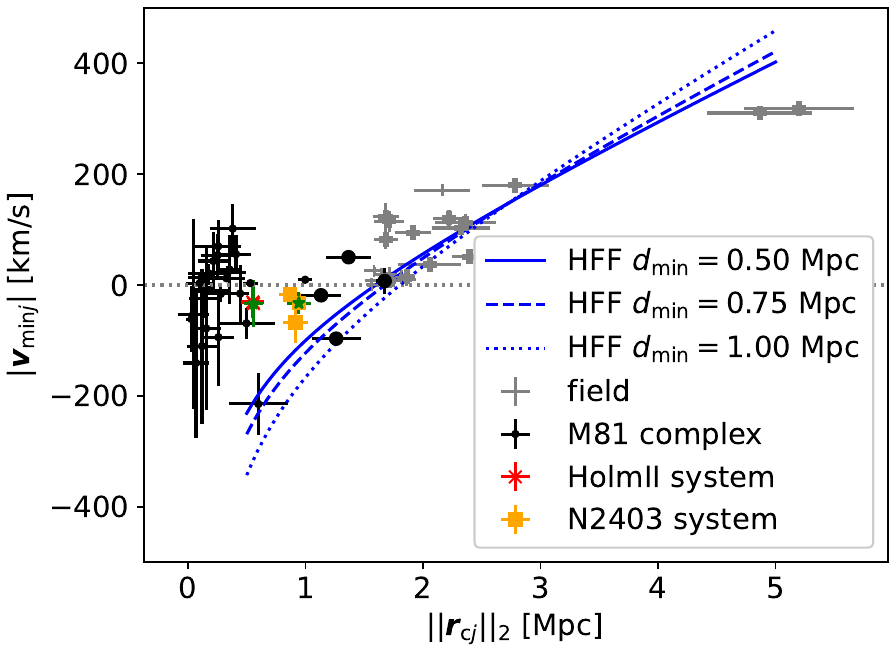} \hspace{3ex}
\includegraphics[width=0.4\linewidth]{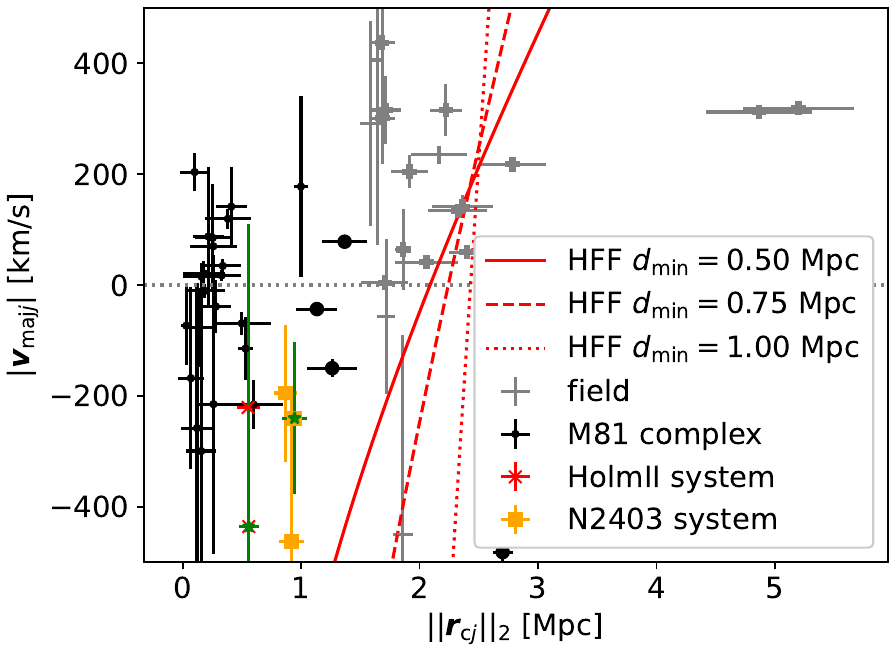} \\[1ex]
\includegraphics[width=0.4\linewidth]{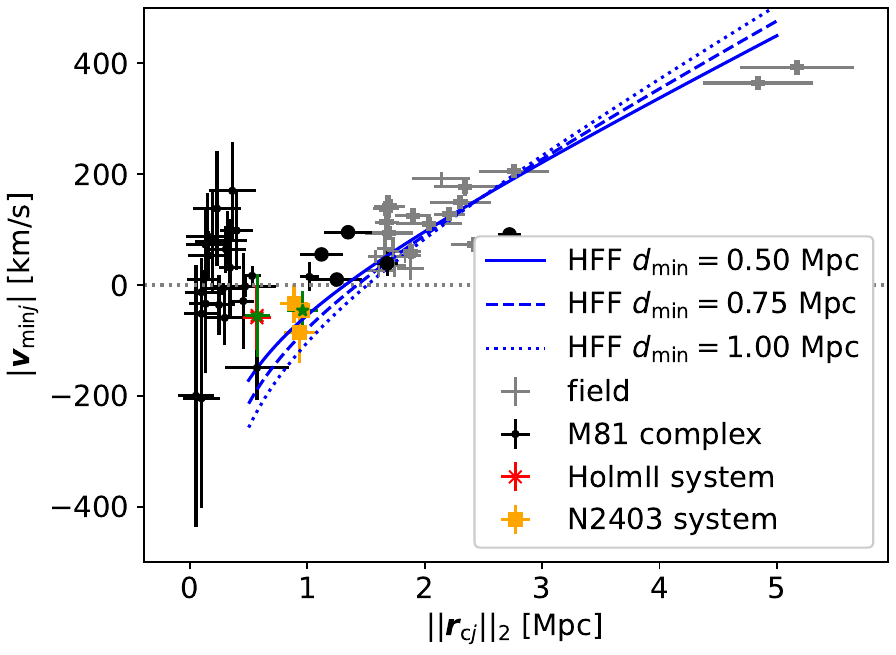} \hspace{3ex}
\includegraphics[width=0.4\linewidth]{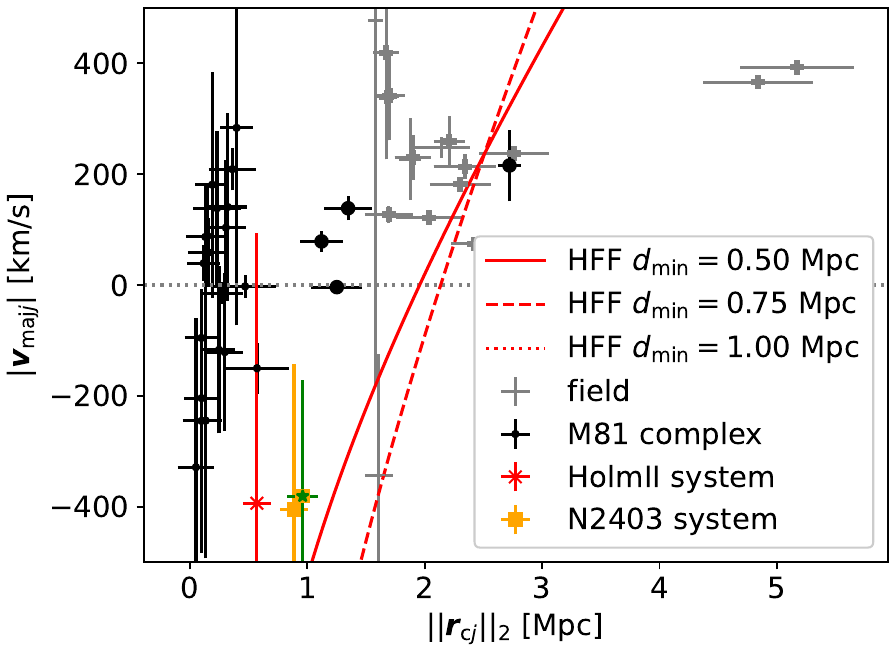} \\[1ex]
\includegraphics[width=0.4\linewidth]{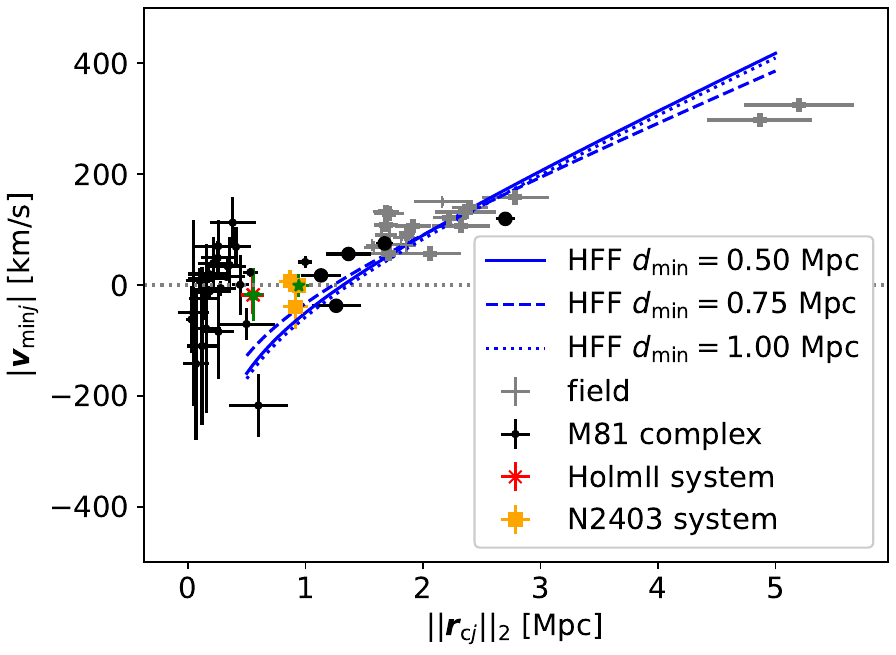} \hspace{3ex}
\includegraphics[width=0.4\linewidth]{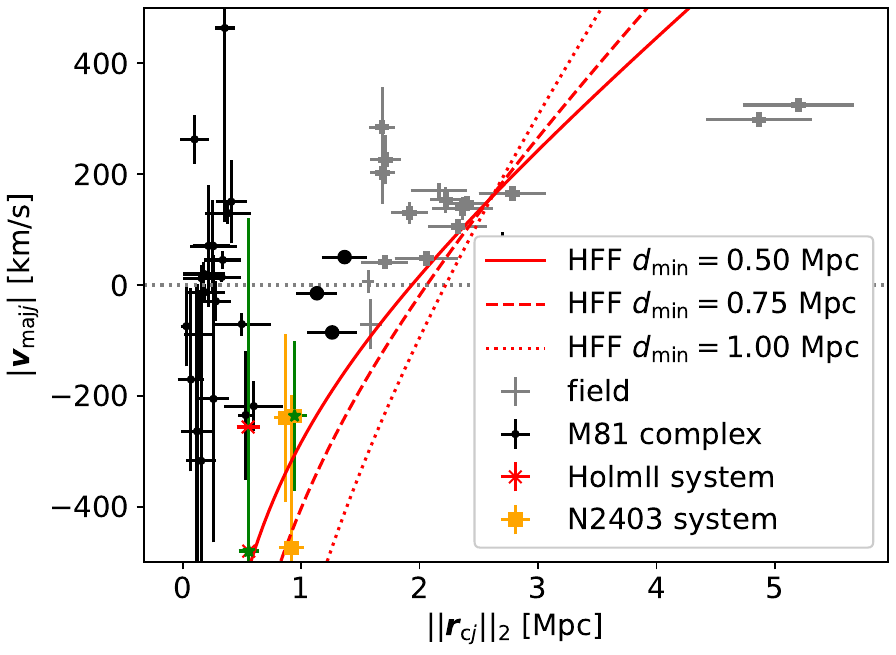}
\caption{\it{Dependence of the Hubble flow fit on the minimum distance from the barycentre for the minor (left column) and major infall models (right column). The first and third rows have M~81 as the centre of mass, the second and fourth row use the barycentre located on the connection line between M~81 and M~82, the first and second row use $v_\mathrm{hel}$, the third and fourth row use $v_\mathrm{LG}$. For the major infall model, velocities up to a maximum value of 700~km/s were included in the fit.}}
\label{fig:dmin_impact_all}
\end{figure*}

\subsection{Impact of parameter dependencies and assumptions}
\label{sec:impact_parameter}

We systematically show the impact of the velocity corrections, the choice of the barycentre, and data cuts on the Hubble flow fit.
The Hubble diagrams in Fig.~\ref{fig:dmin_impact_all} are generated for the four combinations, from top to bottom: (1) M~81 as the barycentre and $v_\mathrm{hel}$ as the basis to calculate the infall velocities, (2) the point on the connection line between M~81 and M~82 as the barycentre and $v_\mathrm{hel}$ for the calculation of the infall velocities, (3) M~81 as the barycentre and $v_\mathrm{LG}$ to determine the infall velocities, (4) the point on the connection line between M~81 and M~82 as the barycentre and $v_\mathrm{LG}$ to calculate the infall velocities.
For each of these four options, the minor or major infall model is used to calculate the infall velocities, as shown on the left side of Fig.~\ref{fig:dmin_impact_all} in blue and on the right side in red, respectively. 
The uncertainties in the three-dimensional distance between all galaxies and the barycentre, $\Vert \boldsymbol{r}_{\mathrm{c}j}\Vert_2$, and in the infall velocities, $\left| \boldsymbol{v}_{\mathrm{inf}j} \right|$, are obtained as detailed in Sect.~\ref{sec:fitting_function}, based on 10,000 samples. 
Increasing the amount of samples to 100,000 does not change the size of the error bars significantly.

In addition, we also investigate variations in the minimum distance from the barycentre, $d_\mathrm{min}$ to start the Hubble flow fit. 
\cite{bib:Karachentsev2006} chose 0.5~Mpc as fiducial value.
We additionally show Hubble flow fits starting at 0.75~Mpc and 1~Mpc distance to the barycentre. 
Due to the instability of the major infall model, we also test the impact of upper bounds on the major infall velocity, using an upper absolute value of 700~km/s and 500~km/s. 

All Hubble flow fits to the data given the uncertainties as calculated by the sampling are obtained by total least squares using a singular value decomposition.  
This is possible because $H_0$ and $M$ are both larger than zero and Eq.~\eqref{eq:HFF} is a linear function in $H_0$ and $\sqrt{M}$ as parameters.
The lines (solid, dashed, dotted) in Fig.~\ref{fig:dmin_impact_all} show the resulting Hubble flow fits for the varying $d_\mathrm{min}$. 
The corresponding best-fit values for $\kappa H_0$ and $M$ are listed in Table~\ref{tab:HFF_svd_results} in an appendix. 
From the figure, it is already clear that the minor infall model yields very robust Hubble flow fits, irrespective of $d_\mathrm{min}$. 
We also note that the variance around the Hubble flow is decreased when using the refined barycentre position and velocity between M~81 and M~82 and additionally applying velocity corrections for our motion in the Local Group. 
The same trends can also be read off the plots for the major infall model, however, with much less robust results due to a diverging denominator in Eq.~\eqref{eq:v_maj} when galaxies are sub-optimally aligned with the barycentre such that the denominator becomes small. 
Table~\ref{tab:HFF_svd_results} even lists unphysical, negative values for $\kappa H_0$ and unrealistically large masses $M$. 
To further investigate the robustness of the major infall model, we studied the impact of different cuts on $\Vert|\boldsymbol{v}_{\mathrm{inf}j}\Vert_2$, as shown in the central and bottom tables of Table~\ref{tab:HFF_svd_results}.
Since both velocity cuts, 700~km/s and 500~km/s, lead to overestimated $H_0$- and $M$-values, the larger spread of the major infall velocities compared to the minor infall velocities causes the less stable Hubble flow fit.
 {From these results, we also conclude that the major infall model, except for the unphysical negative $H_0$-values, can be considered as an upper bound on $\kappa H_0$ and $M$. Establishing the minor infall model as a lower bound for both parameters is motivated from \cite{bib:Benisty2025}: The Illustris TNG50 simulation analysed in their Fig.~3 shows that the starting velocity for the Hubble flow fit is above the true one and the slope for the minor infall model is smaller than the one of the simulated velocities. This result found in mock data seems to be valid for observations based on the results shown in Fig.~\ref{fig:HFF_posterior}.}

Given the amount of at least 20 galaxies available to perform the Hubble flow fits, the resulting parameters are expected to differ only slightly from the ones obtained by ordinary least squares.
Only allowing for shifts in velocity direction, the latter also yields an increased stability compared to the de-regularising total least squares, which allows to shift the Hubble flow fit in distance and velocity directions.
Therefore, our best-fit Hubble flow fit parameters are determined by ordinary least squares in the next section.

\subsection{Best-fit Hubble flow fit}
\label{sec:best-fit}

Our final Hubble flow fit is performed using the barycentre between M~81 and M~82 and accounting for velocity corrections in the Local Group. 
In addition, we enforce $\Vert \boldsymbol{v}_{\mathrm{inf}j} \Vert_2 \in \left[-500, 500\right]$~km/s.
The uncertainties on the distances to the barycentre and infall velocities are again determined as detailed in Sect.~\ref{sec:fitting_function}. 
Galaxies with uncertainties larger than 10\% on the infall velocity and 100\% on the distance to the barycentre are omitted. 

To test the impact of the embedding environment, we perform two fits for each infall model, one including the last two galaxies around 5~Mpc distance from M~81, choosing $d_\mathrm{max}=7$~Mpc, one excluding those galaxies, $d_\mathrm{max}=4$~Mpc. Within these boundaries, there are five satellite galaxies that have $\theta_1 < 0$ for M~81 as their major disturber (black large points in Fig.~\ref{fig:dmin_impact_all}).

\begin{figure}[t!]
    \centering
\includegraphics[width=0.98\linewidth]{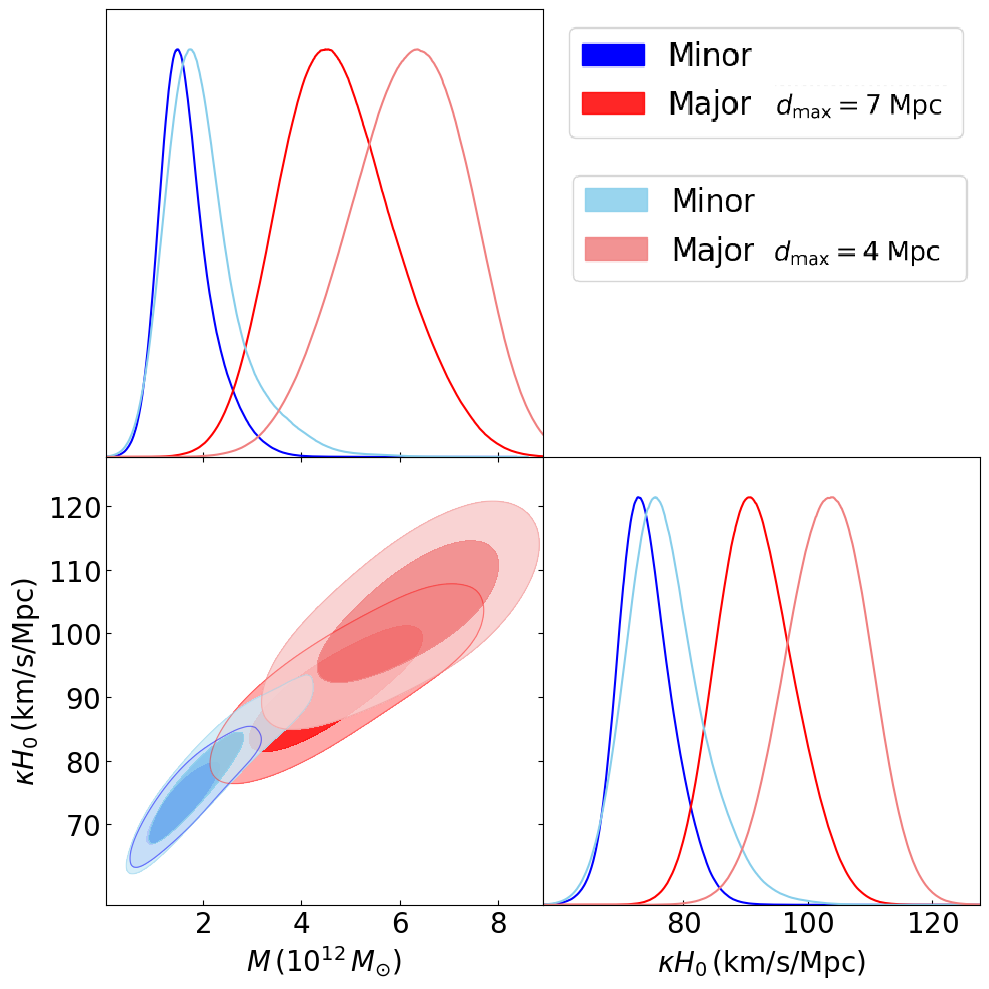}
\begin{tabular}{|l|c|c|c|c|}
\hline
Model & $d_\mathrm{max}$ & $M$ & $\kappa H_0$ & $H_0$ ($\kappa = 1.4$) \\
 & $\left[\mathrm{Mpc} \right]$ & $\left[ 10^{12} M_\odot\right]$ & $\left[ \mathrm{km/s/Mpc}\right]$ & $\left[ \mathrm{km/s/Mpc}\right]$ \\
\hline
Minor & 4 & $2.37 \pm 0.91$ & $81.4 \pm 6.6$ & $58.1 \pm 4.7$ \\
Major & 4 & $6.42 \pm 1.08$ & $108.5 \pm 6.3$ & $77.5 \pm 4.5$ \\
\hline
Minor & 7 & $1.72 \pm 0.54$ & $75.9 \pm 4.4$ & $54.2 \pm 3.1$ \\
Major & 7 & $5.04 \pm 1.20$ & $97.1 \pm 6.3$ & $69.4 \pm 4.5$ \\
\hline
Overlap & 7 & $2.28\pm0.49$ & - & $62.6 \pm 5.4$
\\
\hline 
\end{tabular}
\caption{\label{fig:HFF_posterior}\it{Confidence intervals on $\kappa H_0$ and $M$ for the Hubble flow fit using the minor (blue) and major infall models (red) with respect to the barycentre between M~81 and M~82, a starting distance for the fit of $d_\mathrm{min}=0.5$~Mpc and the two maximum distances of galaxies included in the fit of $d_\mathrm{max}=4$~Mpc (omitting the two farthest galaxies) and $d_\mathrm{max}=7$~Mpc (including all galaxies in the data set). To obtain a robust fit, galaxies with $\Vert \boldsymbol{v}_{\mathrm{maj}j} \Vert_2 > 500$~km/s were excluded. The ``overlap" model yields the best-fit parameter values from the 2-$\sigma$ regions where the minor and major infall models coincide.}}
\end{figure}

We fit Eq.~\eqref{eq:HFF}, the model-based infall velocity $|\boldsymbol{v}^\mathrm{(mod)}_{\mathrm{inf}j}|$, to the infall velocity inferred from the measurements, $|\boldsymbol{v}^\mathrm{(obs)}_{\mathrm{inf}j}|$, via a $\chi^2$-optimisation with:
\begin{equation}
     \chi^2 = -\sum_{j=1}^{n_\mathrm{g}}{\frac{\left( |\boldsymbol{v}^\mathrm{(mod)}_{\mathrm{inf}j}|- |\boldsymbol{v}^\mathrm{(obs)}_{\mathrm{inf}j}|\right)^2}{2\sigma_j^2}} \;,
\end{equation}
in which $\sigma_j$ is given by adding the squared uncertainties in the infall model velocity and the linear propagation of uncertainty of the distance uncertainty: 
\begin{equation}
\sigma_j^2= \left( \Delta |\boldsymbol{v}^{\mathrm{(obs)}}_{\mathrm{inf}j}| \right)^2 + \left( \frac{\partial |\boldsymbol{v}^{\mathrm{(mod)}}_{\mathrm{inf}j}|}{\partial r_{\mathrm{c}j}}\Delta r_{\mathrm{c}j}\right)^2  \;.
\end{equation}
To constrain the best-fit values for $\kappa H_0$ and $M$, a nested sampling is used with the \texttt{PolyChord} package~\citep{bib:Handley2015}. 
We use a uniform prior for $\kappa H_0 \in \left[10,200\right]$~km/s/Mpc and $M \in \left[0.01,8\right] \times 10^{12}~M_\odot$. 
Fig.~\ref{fig:HFF_posterior} is plotted with the \texttt{getdist} package \citep{bib:Lewis2025}, and shows the resulting confidence regions for the minor (blue) and the major infall models (red), once for $d_\mathrm{max}=4$~Mpc (pale regions) and once for $d_\mathrm{max}=7$~Mpc (intensely coloured regions). 
The table underneath the corner plot lists the best-fit values for $\kappa H_0$ and $M$ and their 1-$\sigma$ confidence intervals, as well as an estimate for $H_0$ when fixing $\kappa=1.4$, as assumed in \cite{bib:Penarrubia2014}. 
Moreover, we report the best-fit values for $M$ and $H_0$ in which minor and major infall models coincide within their 2-$\sigma$ confidence bounds, denoted as ``overlap" model.

\begin{figure*}
    \centering
\includegraphics[width=0.58\linewidth]{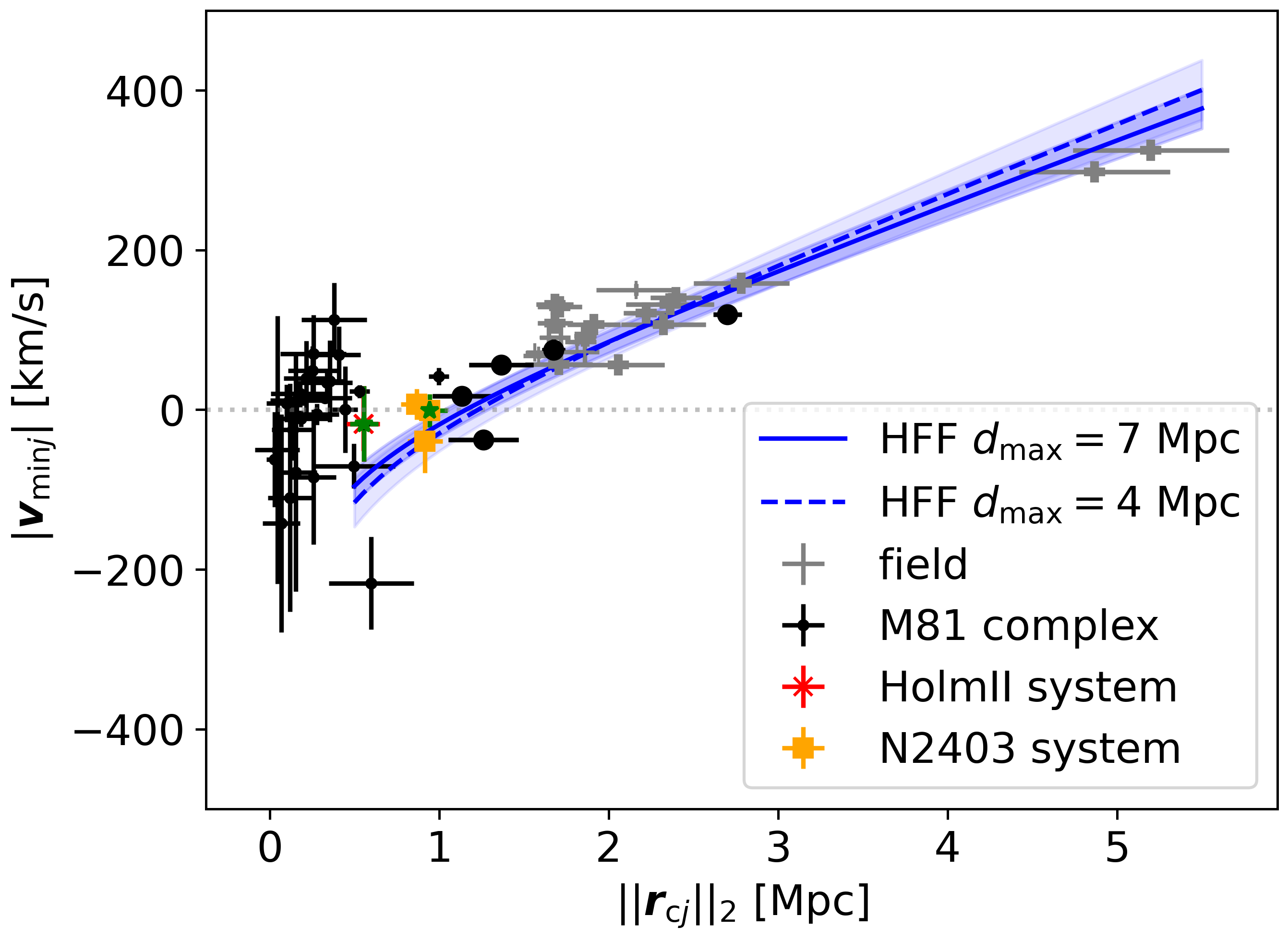} \hfill
\includegraphics[width=0.58\linewidth]{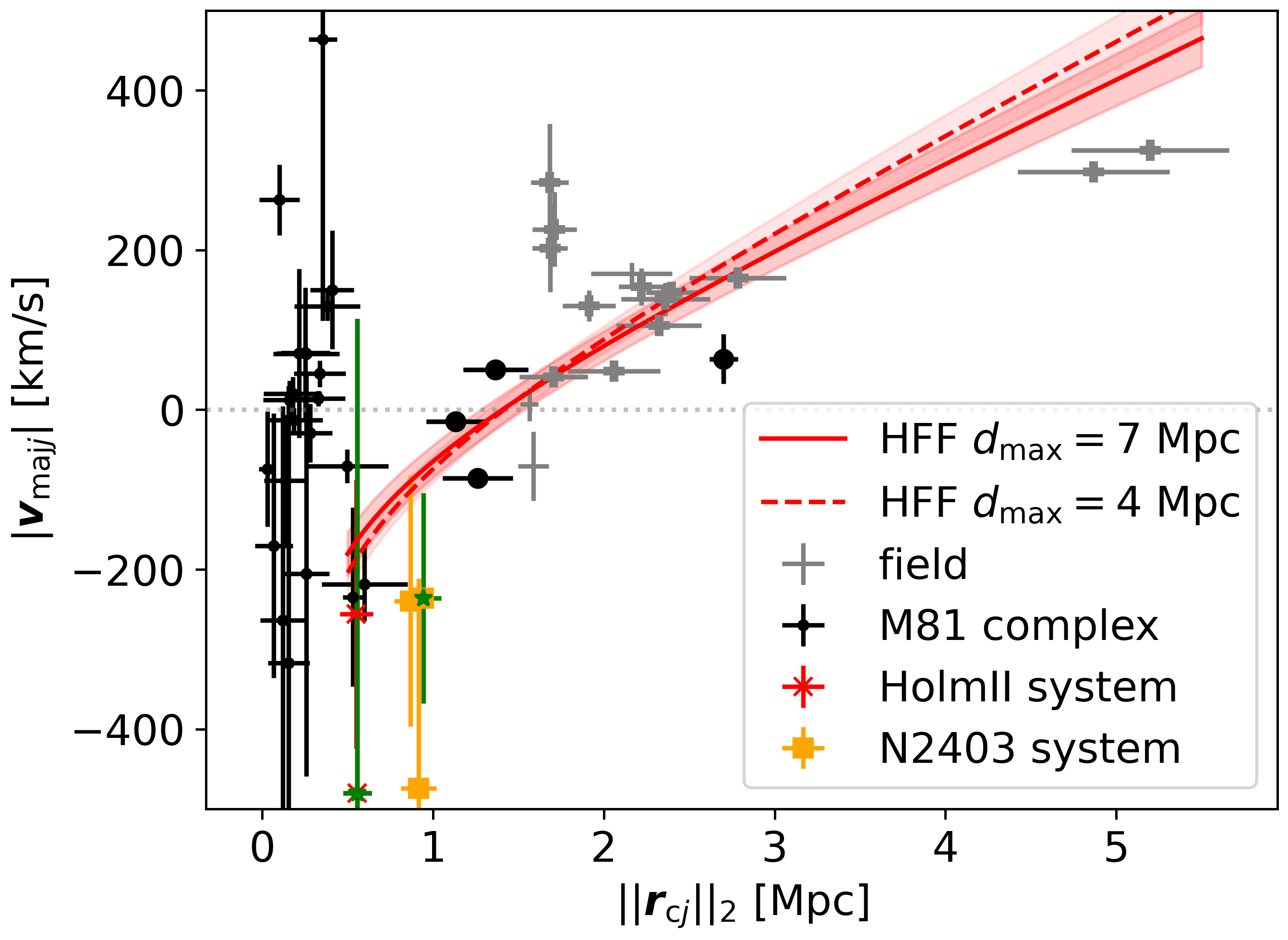}
\caption{\it{Hubble flow fits (Eq.~\ref{eq:HFF}) for the minor (top) and major infall models (bottom) with a starting distance for the fit of $d_\mathrm{min}=0.5$~Mpc and a maximum distance of galaxies included in the fit of $d_\mathrm{max}=7$~Mpc (solid lines) and $d_\mathrm{max}=4$~Mpc (dashed lines). The 1-$\sigma$ confidence bounds around both fits are based on linear propagation of uncertainties for the standard deviations of $\kappa H_0$ and $M$ as given in the table below Fig.~\ref{fig:HFF_posterior}.
}}
\label{fig:HFF_errors}
\end{figure*}

From the table, we read off that excluding the outmost two galaxies from the fit, the best-fit $M$ and $\kappa H_0$ are increased compared to the fit including these galaxies. 
Given that these two galaxies, DDO~87 and UGC05423, both have $\Theta_1 < 0$ for their major disturbers, it is likely that these galaxies belong to the Hubble flow of the M~81 complex. 
While one could choose a different $\kappa$ to account for the varying $d_\mathrm{max}$ when determining $H_0$, we will see in Sect.~\ref{sec:discussion} that the increased mass, particularly for the major infall model, becomes inconsistent with other mass estimation approaches, hinting at the necessity to include these galaxies in the Hubble flow fit. 

For the quality of fit, Fig.~\ref{fig:HFF_errors} shows the Hubble‐flow fits (Eq.~\ref{eq:HFF}) for the minor (top panel) and major (bottom panel) infall models for the 58 galaxies with TRGB-based distances in our data set. 
The fits are shown for two different maximum distances: $d_\mathrm{max}=7$~Mpc (solid lines) and $d_\mathrm{max}=4$~Mpc (dashed lines). 
The shaded regions around the fits indicate 1-$\sigma$ credible intervals, obtained via linear propagation of uncertainties for the standard deviations of $\kappa H_0$ and $M$ listed in Table~\ref{fig:HFF_posterior}.

\section{Dynamical Mass Estimates}
\label{sec:dynamical_mass}

In order to investigate the consistency of the Hubble flow fit and the mass inferred from it, we compare our results obtained in Sect.~\ref{sec:Hubble_flow} with masses inferred from the virial theorem and the projected-mass-method.

\subsection{Hubble-flow-fit mass}

As listed in the table below Fig.~\ref{fig:HFF_posterior}, the minor infall model yields a low mass and Hubble Constant, while the major infall model obtained systematically higher values for $M$ and $H_0$. 
However, the inferred parameter values overlap within their 2-$\sigma$ confidence bounds. 
From the overlapping confidence regions, we infer a common mass of $M = \left(2.28\pm0.49\right)\times 10^{12}\,M_{\odot}$, and $H_0 = \left(62.6 \pm 5.4\right)$~km/s/Mpc. 
Our best-fit $H_0$-values are thus in agreement with \citet{bib:Planck2018} within 1$\sigma$, yet it is also in agreement with \citet{bib:Riess2022} within 2$\sigma$. 

\subsection{Projected mass}

As a second mass estimate, we determine the projected mass, $M_{\rm proj}$, following the formalism of \citet{bib:Bahcall1981,bib:Evans2011}:
\begin{equation}
M_{\rm proj} = \frac{\alpha_\mathrm{proj}}{G} \frac{1}{n_\mathrm{g}} \sum_{j=1}^{n_\mathrm{g}} v_{j}^2 R_{\mathrm{c}j} \;,
\end{equation}
in which each galaxy's line-of-sight velocity is given as $v_j$ and $R_{\mathrm{c}j} \equiv r_\mathrm{c}\theta_{\mathrm{c}j}$ is the physical, projected two-dimensional distance between galaxy~$j$ and the barycentre. 
As usual, $G$ is the gravitational constant and $\alpha_\mathrm{proj}$ is a geometric factor that encapsulates projection and anisotropy effects. 
For an isotropic velocity distribution, we adopt $\alpha_\mathrm{proj} = 16/\pi$. 
Inserting the $v_\mathrm{LG}$ as velocities for all $n_\mathrm{g}=58$ galaxies having TRGB-based distances and the position of the barycentre between M~81 and M~82 as $r_\mathrm{c}$, the mass of the M~81 complex becomes $M_{\rm proj} = (2.74 \pm 0.36)\times 10^{12}\,M_\odot$.
Its uncertainty is inferred from a linear propagation of uncertainties from the measurement uncertainties, see Table~\ref{tab:M81-group}.

\begin{figure*}[t!]
    \centering
    \includegraphics[width=0.85\linewidth]{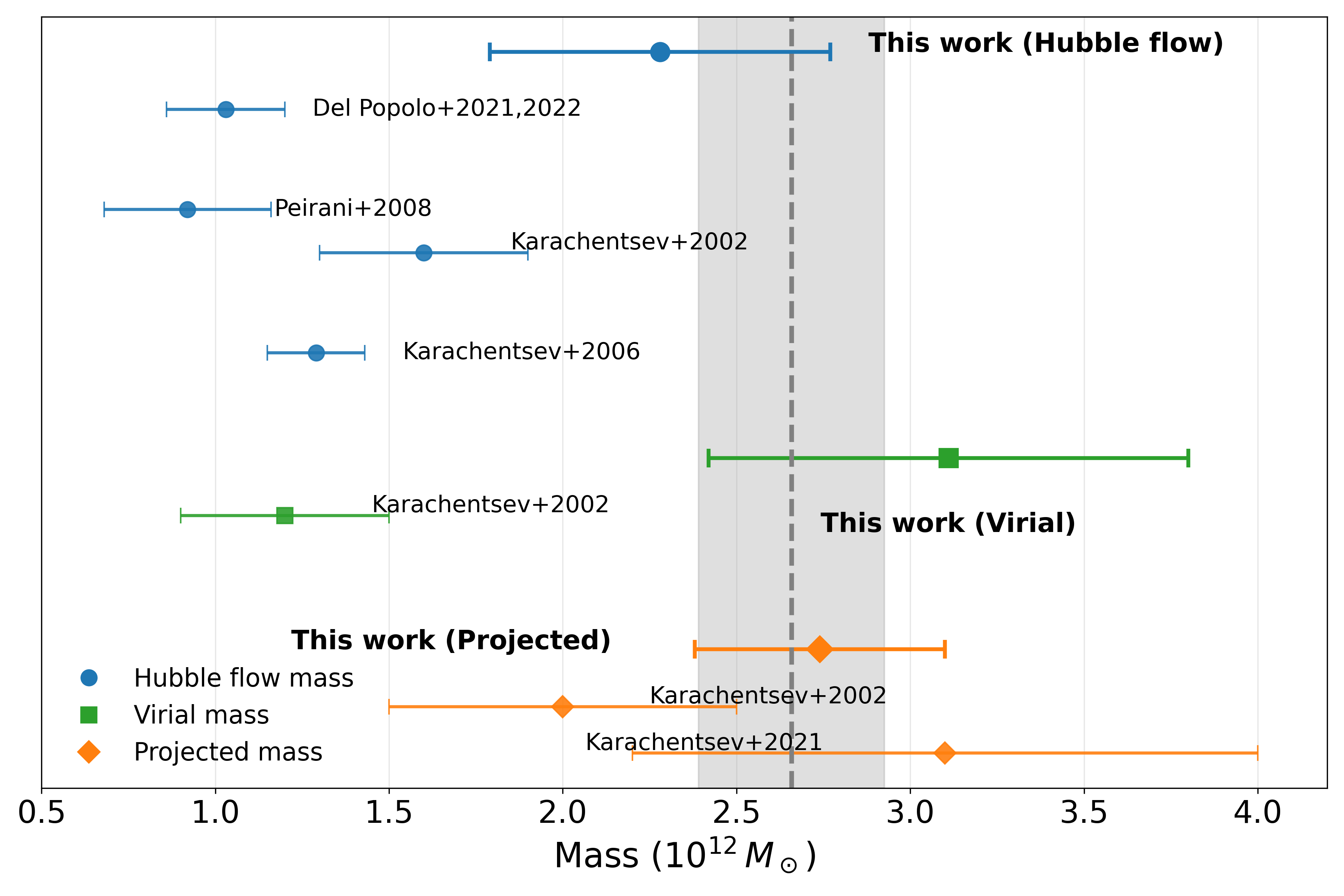}
\caption{\textit{Comparison of M~81-group mass estimates of this work using the Hubble Flow, virial, and projected mass methods (blue circles, green squares, and orange diamonds, respectively), together with mass estimates from the literature.
The grey vertical dashed line and shaded band indicate the weighted mean and 1-$\sigma$ uncertainty of the three estimates of this work. For comparison, literature values are shown from Hubble Flow analyses by \cite{bib:DelPopolo2021,bib:DelPopolo2022}, \cite{bib:Peirani2008}, and \cite{bib:Karachentsev2002, bib:Karachentsev2006};
virial and projected pass estimates from \cite{bib:Karachentsev2002};
and the most recent projected mass by \cite{bib:Karachentsev2021}.}}
    \label{fig:masses_litrature}
\end{figure*}

\subsection{Virial mass}

This estimator assumes that $n_\mathrm{vir}$ satellite galaxies act as test particles in virial equilibrium in a common potential, such that $M_{\rm vir}$ reflects the total gravitationally bound mass within the region sampled by the tracers:
\begin{equation}  
M_{\text{vir}} = \alpha_\mathrm{vir} \frac{\sigma_\mathrm{v}^2 r_\mathrm{G}}{G} \;, \quad \text{with}  \quad r_\mathrm{G} = n_\mathrm{vir}\left(\sum_{j=1}^{n_\mathrm{vir}}{\frac{1}{r_{0j}}}\right)^{-1} \;,
\label{eq:virial_mass}  
\end{equation}
in which $\sigma_\mathrm{v}$ denotes the velocity dispersion of the $n_\mathrm{vir}$ galaxies and $r_{0j}$ is the three-dimensional distance between the centre of the virial volume and galaxy~$j$ \citep{bib:Bahcall1981,bib:Heisler1985,bib:Evans2011,bib:Tully2015a}.
As usual, $G$ is the gravitational constant. The dimensionless coefficient $\alpha_\mathrm{vir}$ represents a geometric form factor accounting for an underlying dark-matter distribution and velocity anisotropies. 
For instance, $\alpha_\mathrm{vir} = 3$ corresponds to isotropic orbits in an isothermal sphere, $\alpha_\mathrm{vir} = 2.6$ to isotropic velocities in a Navarro-Frenk-White profile, and $\alpha_\mathrm{vir} = 2.4$ to specific anisotropy models~\citep{bib:Limber1960,bib:Bahcall1981}.

Modelling the M~81 complex with a single, spherically symmetric potential is a good approximation only for radii much larger than the virial radius.
Since the core is better approximated as a binary system with M~81 and M~82 as the most massive galaxies, see also Sect.~\ref{sec:barycentre}, we generalise the virial mass estimate to a binary system according to \cite{Diaz:2014kqa} and \cite{bib:Benisty2024}.
To do so, we replace $r_{0j}$ in Eq.~\eqref{eq:virial_mass} by the mass-weighted average:
\begin{equation}
r_\mathrm{G} = n_\mathrm{vir}\left(\sum_{j=1}^{n_\mathrm{vir}} \frac{m_{\mathrm{M81}}}{r_{\mathrm{M81} j}} + \frac{ 1 - m_\mathrm{M81}}{r_{\mathrm{M82} j}}\right)^{-1},
\end{equation}
in which each satellite galaxy~$j$ has a distance $r_{\mathrm{M81}j}$ to the centre of M~81 and $r_{\mathrm{M82}j}$ to M~82.
The mass ratio is $m_\mathrm{M81}\equiv M_{\mathrm{M81}}/(M_{\mathrm{M81}}+M_{\mathrm{M82}})$ with the masses of M~81 and M~82 to be $M_{\mathrm{M81}}$ and $M_{\mathrm{M82}}$, based on their luminosity as in Sect.~\ref{sec:barycentre}.
Then, the velocity dispersion $\sigma_\mathrm{v}$ is determined with respect to the barycentre between M~81 and M~82 from Sect.~\ref{sec:barycentre}.
To encompass uncertainties in the dark matter distribution, we adopt a broad uniform prior of $\alpha_\mathrm{vir} \in [2, 4]$. 
For the $n_\mathrm{vir} = 36$ member galaxies bound to the M~81 complex, we estimate $\sigma_\mathrm{v} = (106 \pm 4)$~km/s  {from the stable minor infall model velocities multiplied by $\sqrt{3}$, which is in good agreement with the previous best-fit value of $\sigma_\mathrm{v}=102$~km/s obtained in \cite{bib:Wagner2025}. Moreover,} $r_\mathrm{G} = (0.41 \pm 0.04)$~Mpc, which is close to $d_\mathrm{min} = 0.5$~Mpc as the starting radius for the Hubble flow fit, from the data in Table~\ref{tab:M81-group} to arrive at $M_{\rm vir} = (3.11 \pm 0.69)\times 10^{12}~M_\odot$. 
 


\section{Conclusion}
\label{sec:discussion}

We employed precise TRGB-based distances for nine previously unmeasured or imprecisely measured galaxies to determine the dynamics of M~81 complex. 
This expanded dataset, now comprising of 58 galaxies with high-precision TRGB-based distances, has allowed us to perform a robust dynamical analysis of the group's core and its embedding into the local Hubble flow. 

Despite the improved amount and precision of the data, finding an alignment of the M~81 complex with the supergalactic SGX-SGY-plane still depends on a few galaxies between the second turnaround radius of ca.~0.23~Mpc and the virial radius of ca.~0.41~Mpc. 
The alignment, however, becomes robust beyond the virial radius, which formed the outer limit of the study by \cite{bib:Mueller2024}. 
Our study in Sect.~\ref{sec:embedding_environment} revealed the filamentary structure around the M~81 group in a volume of about 5~Mpc radius. 
Hence, the filaments found are on a scale larger than the satellite plane of \cite{bib:Mueller2024} and on a smaller scale than the filaments of \cite{bib:Raj2024}.

Within this environment, our data set contains a Hubble flow of more than 30 galaxies for which we perform a Hubble flow fit.
Under the assumption that light traces mass, we demonstrate that this Hubble flow fit is best performed from the barycentre of the group determined as the point on the connection line between the two most luminous galaxies in the group, M~81 and M~82.
Incorporating more galaxies, like NGC~3077 and N~2976, does not change the position or line-of-sight velocity of the barycentre further for the given measurement precision.
Moreover, it is important to account for the observers' motion in the Local Group to alleviate the variance in the infall velocities of the satellite galaxies onto the barycentre of the M~81 complex. 

We also investigate the impact of the satellite selection for the Hubble flow fit.
Both models systematically increase their best-fit $M$ and $H_0$ values when restricting the fitting region to a maximum distance of 4~Mpc from the barycentre. 
Identifying the most distant galaxies on the Hubble flow is thus not only important to obtain an accurate value for $H_0$, but also for $M$. 
Varying the starting distance for the Hubble flow fit, we find that the minor infall model yields more robust mass and $H_0$ values for different starting and ending points of the fit than the major infall model, which is highly sensitive to these choices. 
In general, analogously to \cite{bib:Wagner2025}, we find that the major infall model shows a larger dispersion in the infall velocities and that the minor and major infall models can be considered as lower and upper bounds for the true Hubble flow fit parameters. 
We therefore determine the Hubble flow fit parameters $M_\mathrm{HFF} = (2.28\pm0.49)\times 10^{12} M_\odot$ and $H_0 = (62.6 \pm 5.4)$~km/s/Mpc from the overlapping 2-$\sigma$ bounds on both infall models. 
For the latter, we set $\kappa=1.4$ like \cite{bib:Penarrubia2014}, fixing the degeneracy between the background embedding and $H_0$ in the Hubble flow formula. 
Choosing smaller $\kappa$, $H_0$ can be larger, shifting the agreement from \cite{bib:Planck2018} to \cite{bib:Riess2022}.
Our $H_0$ value with an imprecision of about 8\% is not precise enough to alleviate the Hubble tension, but it is consistent with prior works inferring $H_0$ in the Local-Universe, using TRGB-based distances, see, for instance, \cite{bib:Kim2020} for a similar result of a Hubble flow fit in the Virgo cluster and \cite{bib:DESI2025} and references therein for a more details on other $H_0$-probes. 
\cite{bib:Tully2023} also discuss the TRGB-zero-point calibrations and current disputes over it, potentially causing the discrepancy to Cepheid-based distances in the Local Volume. 

Comparing the Hubble flow fit $M$ of the entire M~81 complex with the virial and projected mass estimates based on our improved data set, a more consistent picture emerges compared to prior work from the literature, as summarised in Fig.~\ref{fig:masses_litrature}. 
Analyses by \cite{bib:DelPopolo2021,bib:DelPopolo2022}, and \cite{bib:Peirani2008} obtained masses from a Hubble flow fit close to $10^{12}~M_{\odot}$, while slightly higher values were reported by \cite{bib:Karachentsev2002} and \cite{bib:Karachentsev2006}. 
Using alternative approaches, such as the virial theorem and projected mass estimators, \cite{bib:Karachentsev2002} and \cite{bib:Karachentsev2021} obtained comparable or higher masses. Overall, the literature suggested that the M~81 system has a total mass of the order of $(1\text{–}2)\times10^{12}~M_{\odot}$, consistent with expectations for a relatively low-mass, nearby galaxy group dominated by a few large spirals.
Yet, previous mass estimates were based on even sparser data sets and only used the minor infall model, which jointly led to systematically lower mass estimates around $10^{12}~M_{\odot}$ than the values obtained in this work, $M_\mathrm{proj} = (2.74 \pm 0.36)~\times~10^{12}~M_\odot$ and $M_\mathrm{vir}=(3.11\pm 0.69)~\times 10^{12}~M_\odot$.
The latter, however, is subject to a very large uncertainty based on the known logarithmic divergence of the harmonic radius $r_\mathrm{G}$ \citep{bib:Bahcall1981}. 
 
The consistency among the three mass estimation techniques in our analysis suggests that the higher mass is not an artefact of a single estimator, but rather reflects a genuine physical difference that may arise from updated kinematics, improved distance measurements, and a more accurate treatment of the system’s gravitational potential. 
The higher inferred masses imply that M~81 may be dynamically more evolved and more strongly bound than earlier studies indicated, with potential implications for its satellite population and the distribution of dark matter within the group.
The already planned HST observations in the SNAP 18070 programme of eight galaxies with poorly constrained distances UGC06451, KKH30, SMDG0956+82, KDG162, Dw1907+63, Dw1940+64, Dw1559+46, and Dw1735+57 in the upcoming C~33 cycle will certainly bring further clarity to check our results in the near future.

\begin{acknowledgements}
DB is supported by the Minerva Stiftung Gesellschaft für die Forschung mbH. This article is based on work from the COST Action CA21136 - "Addressing observational tensions in cosmology with systematics and fundamental physics" (CosmoVerse) and the Cost Action CA23130 - “Bridging high and low energies in search of quantum gravity (BridgeQG), supported by COST (European Cooperation in Science and Technology). 
IK is supported by the Russian Science Foundation grant No 24-12-00277, ``Cosmography of the Local Volume of the Universe within 15
megaparsecs".
\end{acknowledgements}

\bibliography{ref}{}
\bibliographystyle{aasjournal}


\section{Observational data sets}
\label{app:observational_data}

\begin{longtable}[h!]{lrrrrrcrrcccc}
\caption{\textit{M81-group members with their name, coordinates on the sky (J2000), velocity with respect to the heliocentric frame, $v_\mathrm{hel}$, and its uncertainty, $\Delta v_\mathrm{hel}$, magnitude in the $B$-band, their ``major disturber" (MD) and tidal index to their MD, $\Theta_1$, the distance to us as observers, $r$, and the method to determine the distance. The last column indicates if the galaxy was used for the Hubble flow fit (HFF) in Sect.~\ref{sec:best-fit}. Rows are sorted according to their distance to M~81. Galaxies in the data set by \cite{bib:Mueller2024} have a * after their name. \label{tab:M81-group}}}\\   

\hline
  \multicolumn{1}{c}{Name} &
  \multicolumn{1}{c}{$\alpha$} &
  \multicolumn{1}{c}{$\delta$} &
  \multicolumn{1}{c}{$v_\mathrm{hel}$} &
  \multicolumn{1}{c}{$\Delta v_\mathrm{hel}$} &
  \multicolumn{1}{c}{B} &
  \multicolumn{1}{c}{MD} &
   \multicolumn{1}{c}{$\Theta_1$} &
  \multicolumn{1}{c}{$r$} & 
  \multicolumn{1}{c}{Method} &
  \multicolumn{1}{c}{HFF} \\
   \multicolumn{1}{c}{} &
  \multicolumn{1}{c}{$\left[ \mathrm{deg}\right]$} &
  \multicolumn{1}{c}{$\left[ \mathrm{deg}\right]$} &
  \multicolumn{1}{c}{$\left[ \mathrm{km}/\mathrm{s}\right]$} &
  \multicolumn{1}{c}{$\left[ \mathrm{km}/\mathrm{s}\right]$} &
  \multicolumn{1}{c}{$\left[ \mathrm{mag}\right]$} &
  \multicolumn{1}{c}{} &
  \multicolumn{1}{c}{$\left[ \mathrm{deg}\right]$} &
  \multicolumn{1}{c}{$\left[ \mathrm{Mpc}\right]$} &
  \multicolumn{1}{c}{} &
  \multicolumn{1}{c}{} \\
\hline
 \endfirsthead

\hline
  \multicolumn{1}{c}{Name} &
  \multicolumn{1}{c}{$\alpha$} &
  \multicolumn{1}{c}{$\delta$} &
  \multicolumn{1}{c}{$v_\mathrm{hel}$} &
  \multicolumn{1}{c}{$\Delta v_\mathrm{hel}$} &
  \multicolumn{1}{c}{B} &
  \multicolumn{1}{c}{MD} &
   \multicolumn{1}{c}{$\Theta_1$} &
  \multicolumn{1}{c}{$r$} & 
  \multicolumn{1}{c}{Method} &
  \multicolumn{1}{c}{HFF} \\
   \multicolumn{1}{c}{} &
  \multicolumn{1}{c}{$\left[ \mathrm{deg}\right]$} &
  \multicolumn{1}{c}{$\left[ \mathrm{deg}\right]$} &
  \multicolumn{1}{c}{$\left[ \mathrm{km}/\mathrm{s}\right]$} &
  \multicolumn{1}{c}{$\left[ \mathrm{km}/\mathrm{s}\right]$} &
  \multicolumn{1}{c}{$\left[ \mathrm{mag}\right]$} &
  \multicolumn{1}{c}{} &
  \multicolumn{1}{c}{$\left[ \mathrm{deg}\right]$} &
  \multicolumn{1}{c}{$\left[ \mathrm{Mpc}\right]$} &
  \multicolumn{1}{c}{} &
  \multicolumn{1}{c}{} \\
\hline
\endhead

\hline
\endfoot
 
\hline
\endlastfoot
  MESSIER081$^*$	&	148.889583	&	69.066667	&	-38	&	1	&	7.8	&	 M82 	&	2.7	&	3.7	&	 TRGB 	&	\\
  JKB83 	&	148.956667	&	69.332500	&	56	&	10	&	19.7	&	 M81 	&	5.3	&	3.7	&	 mem 	&	\\
  KDG61em 	&	149.281250	&	68.598333	&	116	&	21	&	18.8	&	 M81 	&	4.5	&	3.7	&	 mem 	&	\\
  KDG061$^*$ 	&	149.261250	&	68.591667	&	221	&	3	&	15.2	&	 M81 	&	4	&	3.66	&	 TRGB 	&	\\
  NGC2976$^*$ 	&	146.815000	&	67.913611	&	6	&	4	&	11	&	 M81 	&	3	&	3.66	&	 TRGB 	&	\\
  MESSIER082$^*$ 	&	148.974583	&	69.682500	&	183	&	4	&	9.1	&	 M81 	&	3.1	&	3.61	&	 TRGB 	&	\\
  IKN$^*$ 	&	152.024583	&	68.399167	&	-140	&	64	&	14.5	&	 M81 	&	3	&	3.75	&	 TRGB 	&	\\
  ClumpI 	&	149.338333	&	68.715278	&	-165	&	18	&	19.8	&	 M82 	&	3.2	&	3.6	&	 mem 	&	\\
  ClumpIII 	&	150.168333	&	68.660278	&	-121	&	20	&	19.8	&	 M82 	&	3.1	&	3.6	&	 mem 	&	\\
  KDG064$^*$ 	&	151.757917	&	67.827500	&	-15	&	13	&	15.5	&	 M81 	&	2.8	&	3.75	&	 TRGB 	&	\\
  GARLAND$^*$ 	&	150.925000	&	68.693333	&	43	&	17	&	16.8	&	 N3077 	&	3	&	3.82	&	 TRGB 	&	\\
  F8D1$^*$ 	&	146.196250	&	67.438611	&	-125	&	130	&	16.8	&	 M81 	&	2.6	&	3.75	&	 TRGB 	&	\\
  HIJASS 1021+68 	&	155.250833	&	68.700000	&	46	&	1	&	  	&	 M81 	&	2.5	&	3.7	&	 mem 	&	\\
  HolmIX$^*$ 	&	149.385000	&	69.043056	&	50	&	4	&	14.5	&	 N3077 	&	2.8	&	3.85	&	 TRGB 	&	\\
  NGC3077$^*$ 	&	150.837500	&	68.733889	&	19	&	4	&	10.6	&	 M81 	&	2.3	&	3.85	&	 TRGB 	&	\\
 d0958+66$^*$ 	&	149.702083	&	66.849722	&	90	&	50	&	16	&	 M81 	&	2.1	&	3.82	&	 TRGB 	&	\\
  A0952+69$^*$ 	&	149.370833	&	69.272222	&	99	&	2	&	16.8	&	 M81 	&	1.9	&	3.93	&	 TRGB 	&	\\
 d1028+70$^*$ 	&	157.165417	&	70.233611	&	-114	&	50	&	16.2	&	 M81 	&	1.8	&	3.84	&	 TRGB 	&	\\
 d0944+71 	&	146.143333	&	71.482500	&	-38	&	10	&	16.7	&	 M82 	&	1.8	&	3.47	&	 TRGB 	&	\\
  DDO078$^*$ 	&	156.616250	&	67.656667	&	55	&	10	&	14.3	&	 M81 	&	1.6	&	3.48	&	 TRGB 	&	\\
  IC2574$^*$ 	&	157.093333	&	68.416111	&	43	&	4	&	11.5	&	 M81 	&	1.5	&	3.93	&	 TRGB 	&	\\
  DDO082$^*$ 	&	157.645833	&	70.619444	&	56	&	3	&	13.5	&	 M81 	&	1.5	&	3.93	&	 TRGB 	&	\\
  UGC05497$^*$ 	&	153.201667	&	64.107500	&	150	&	50	&	15.6	&	 M81 	&	1.4	&	3.73	&	 TRGB 	&	\\
  HolmI$^*$ 	&	145.134583	&	71.186389	&	139	&	1	&	13.6	&	 M81 	&	1.3	&	4.02	&	 TRGB 	&	\\
  KDG073 	&	163.237917	&	69.545833	&	116	&	1	&	17.1	&	 M81 	&	1.2	&	3.91	&	 TRGB 	&	\\
  UGC04483 	&	129.262500	&	69.775278	&	156	&	1	&	15	&	 M81 	&	1	&	3.58	&	 TRGB 	&	\\
  BK3N$^*$ 	&	148.452083	&	68.969167	&	-40	&	20	&	18.8	&	 M81 	&	1	&	4.17	&	 TRGB 	&	$\checkmark$\\
  DDO053 	&	128.527083	&	66.179167	&	19	&	2	&	14.6	&	 M81 	&	0.8	&	3.68	&	 TRGB 	&	\\
  KDG052 	&	125.983333	&	71.029444	&	116	&	2	&	16.4	&	 HolmII 	&	2	&	3.42	&	 TRGB 	&	\\
 d0959+68 	&	149.887917	&	68.656944	&	-186	&	44	&	18	&	 M81 	&	0.7	&	4.27	&	 TRGB 	&	$\checkmark$\\
  HolmII 	&	124.766667	&	70.714167	&	157	&	1	&	11.1	&	 M81 	&	0.7	&	3.47	&	 TRGB 	&	\\
  UGC06451 	&	172.193333	&	79.601944	&	50	&	1	&	16.5	&	 M81 	&	0.3	&	3.6	&	 mem 	&	\\
  NGC2366 	&	112.227500	&	69.205278	&	96	&	1	&	11.7	&	 N2403 	&	0.8	&	3.28	&	 TRGB 	&	\\
  DDO44 	&	113.547083	&	66.886111	&	213	&	25	&	15.6	&	 N2403 	&	2.3	&	3.21	&	 TRGB 	&	\\
  NGC2403 	&	114.214167	&	65.599444	&	125	&	17	&	8.8	&	 DDO44 	&	0.2	&	3.19	&	 TRGB 	&	\\
  KKH37 	&	101.940833	&	80.123889	&	11	&	7	&	16.4	&	 M81 	&	0	&	3.44	&	 TRGB 	&	\\
  NGC4236 	&	184.180417	&	69.465556	&	-3	&	1	&	10.1	&	 M81 	&	-0.2	&	4.41	&	 TRGB 	&	$\checkmark$\\
  KKH30 	&	79.425000	&	81.624167	&	-67	&	20	&	17.5	&	 M81 	&	-0.2	&	3.5	&	 txt 	&	\\
  UGC06456 	&	172.002500	&	78.991389	&	-94	&	2	&	14.3	&	 M81 	&	-0.3	&	4.63	&	 TRGB 	&	$\checkmark$\\
  SMDG0956+82 	&	149.054167	&	82.890000	&	-63	&	10	&	18.1	&	 M81 	&	-0.4	&	2.6	&	 NAM 	&	\\
  UGC06757 	&	176.746250	&	61.334722	&	88	&	1	&	16.3	&	 M81 	&	-0.4	&	4.61	&	 TRGB 	&	$\checkmark$\\
  KDG162 	&	188.756667	&	58.385556	&	125	&	1	&	17.7	&	 M81 	&	-0.5	&	2.8	&	 NAM 	&	\\
  CamA 	&	66.315000	&	72.805833	&	-54	&	2	&	14.8	&	 IC342 	&	0.7	&	3.56	&	 TRGB 	&	\\
  NGC1560 	&	68.207917	&	71.881111	&	-36	&	1	&	12.1	&	 IC342 	&	0.8	&	2.99	&	 TRGB 	&	$\checkmark$\\
  CamB 	&	73.278750	&	67.099167	&	78	&	2	&	16.7	&	 IC342 	&	0.7	&	3.5	&	 TRGB 	&	\\
  UGC06541 	&	173.371250	&	49.238056	&	249	&	2	&	14.4	&	 N4736 	&	-0.6	&	4.23	&	 TRGB 	&	$\checkmark$\\
  UGC07298 	&	184.119167	&	52.227222	&	174	&	2	&	16	&	 N4736 	&	-0.4	&	4.19	&	 TRGB 	&	$\checkmark$\\
  NGC3741 	&	174.026667	&	45.285278	&	229	&	2	&	14.4	&	 M81 	&	-0.7	&	3.22	&	 TRGB 	&	\\
  Dw1245+6158 	&	191.454167	&	61.968889	&	68	&	40	&	18.4	&	  	&	  	&	4.72	&	 sbf 	&	\\
  DDO165 	&	196.611667	&	67.704167	&	31	&	1	&	13.3	&	 N4236 	&	-0.6	&	4.83	&	 TRGB 	&	$\checkmark$\\
  NGC4068 	&	181.010000	&	52.588611	&	210	&	2	&	13.2	&	 N4736 	&	-0.5	&	4.39	&	 TRGB 	&	$\checkmark$\\
  UGCA105 	&	78.562917	&	62.580833	&	113	&	22	&	12.1	&	 IC342 	&	0.3	&	3.39	&	 TRGB 	&	\\
  KKH22 	&	56.235833	&	72.064444	&	30	&	10	&	17	&	 IC342 	&	1.3	&	3.12	&	 TRGB 	&	\\
  NGC1569 	&	67.704583	&	64.848056	&	-86	&	3	&	11.8	&	 IC342 	&	1.1	&	3.19	&	 TRGB 	&	\\
  UGCA092 	&	68.001250	&	63.613889	&	-95	&	2	&	15.2	&	 N1569 	&	1.8	&	3.22	&	 TRGB 	&	\\
  UGCA086 	&	59.956250	&	67.125278	&	72	&	5	&	13.5	&	 IC342 	&	1.1	&	2.98	&	 TRGB 	&	\\
  IC0342 	&	56.703750	&	68.095833	&	29	&	1	&	9.4	&	 N1569 	&	-0.1	&	3.28	&	 TRGB 	&	\\
  MCG09-20-131 	&	183.944583	&	52.387500	&	159	&	2	&	15.3	&	 N4736 	&	-0.4	&	4.61	&	 TRGB 	&	$\checkmark$\\
  UGC07242 	&	183.530833	&	66.092222	&	66	&	2	&	14.6	&	 N4605 	&	-0.4	&	5.45	&	 TRGB 	&	$\checkmark$\\
  NGC3738 	&	173.952500	&	54.522778	&	225	&	8	&	11.9	&	 Dw1135+54 	&	1.3	&	5.3	&	 TRGB 	&	$\checkmark$\\
  KK109 	&	176.796667	&	43.671944	&	211	&	2	&	18.1	&	 N4736 	&	-0.3	&	4.51	&	 TRGB 	&	$\checkmark$\\
  KK135 	&	184.894583	&	58.042778	&	142	&	1	&	17.5	&	 N4605 	&	-0.2	&	5.46	&	 TRGB 	&	$\checkmark$\\
  NGC4605 	&	190.001250	&	61.608056	&	151	&	17	&	10.9	&	 M101 	&	-1.1	&	5.55	&	 TRGB 	&	$\checkmark$\\
  UGC04879 	&	139.009167	&	52.840000	&	-25	&	4	&	13.8	&	 M31 	&	-0.7	&	1.37	&	 TRGB 	&	$\checkmark$\\
  Dw1907+63 	&	286.814167	&	63.385000	&	-147	&	10	&	18.3	&	 M31 	&	-1.1	&	2.4	&	 NAM 	&	\\
  NGC6789 	&	289.174167	&	63.972778	&	-140	&	9	&	13.8	&	 M81 	&	-1.3	&	3.55	&	 TRGB 	&	$\checkmark$\\
  Grapes 	&	178.023333	&	54.792222	&	223	&	1	&	18.3	&	 N4258 	&	-0.9	&	5.96	&	 TRGB 	&	$\checkmark$\\
  Dw1940+64 	&	295.230833	&	64.758889	&	-59	&	10	&	18	&	 M81 	&	-1.4	&	3.8	&	 NAM 	&	\\
  Dw1559+46 	&	239.760833	&	46.394444	&	77	&	1	&	17.2	&	 M81 	&	-1.4	&	3.4	&	 NAM 	&	\\
  Dw1735+57 	&	263.894167	&	57.813056	&	43	&	1	&	17.1	&	 M81 	&	-1.6	&	4.6	&	 NAM 	&	\\
  DDO 087 	&	162.402083	&	65.530556	&	340	&	6	&	15.2	&	 N2787 	&	-1.4	&	8.5	&	 TRGB 	&	$\checkmark$\\
  UGC05423 	&	151.377500	&	70.364444	&	348	&	1	&	14.4	&	 CKT1009+70 	&	-0.5	&	8.87	&	 TRGB 	&	$\checkmark$\\
\end{longtable}

\begin{longtable}[h!]{lrrr}
\caption{\textit{Galaxy positions used to analyse the embedding of the M~81 group into the surroundings in Sect.~\ref{sec:embedding_environment}. Rows are sorted according to their distance to us. The first part of the table lists the 48 galaxies from Table~\ref{tab:M81-group} used in this analysis, the second part lists 55 galaxies in the environment. As we only employ the galaxy locations, we do not list velocities here. \label{tab:M81-environment}}}\\   

\hline
  \multicolumn{1}{c}{Name} &
  \multicolumn{1}{c}{$\alpha$} &
  \multicolumn{1}{c}{$\delta$} &
  \multicolumn{1}{c}{$r$} \\
  \multicolumn{1}{c}{} &
  \multicolumn{1}{c}{$\left[ \mathrm{deg}\right]$} &
  \multicolumn{1}{c}{$\left[ \mathrm{deg}\right]$} &
  \multicolumn{1}{c}{$\left[ \mathrm{Mpc}\right]$} \\
\hline
 \endfirsthead

\hline
  \multicolumn{1}{c}{Name} &
  \multicolumn{1}{c}{$\alpha$} &
  \multicolumn{1}{c}{$\delta$} &
  \multicolumn{1}{c}{$r$} \\
  \multicolumn{1}{c}{} &
  \multicolumn{1}{c}{$\left[ \mathrm{deg}\right]$} &
  \multicolumn{1}{c}{$\left[ \mathrm{deg}\right]$} &
  \multicolumn{1}{c}{$\left[ \mathrm{Mpc}\right]$} \\
\hline
\endhead

\hline
\endfoot
 
\hline
\endlastfoot

UGCA086	&	59.956250	&	67.125278	&	2.98	\\
NGC1560	&	68.207917	&	71.881111	&	2.99	\\
KKH22	&	56.235833	&	72.064444	&	3.12	\\
NGC1569	&	67.704583	&	64.848056	&	3.19	\\
NGC2403	&	114.214167	&	65.599444	&	3.19	\\
DDO44	&	113.547083	&	66.886111	&	3.21	\\
NGC3741	&	174.026667	&	45.285278	&	3.22	\\
UGCA092	&	68.001250	&	63.613889	&	3.22	\\
NGC2366	&	112.227500	&	69.205278	&	3.28	\\
UGCA105	&	78.562917	&	62.580833	&	3.39	\\
KDG052	&	125.983333	&	71.029444	&	3.42	\\
KKH37	&	101.940833	&	80.123889	&	3.44	\\
d0944+71	&	146.143333	&	71.482500	&	3.47	\\
HolmII	&	124.766667	&	70.714167	&	3.47	\\
DDO0 78	&	156.616250	&	67.656667	&	3.48	\\
CamB	&	73.278750	&	67.099167	&	3.5	\\
CamA	&	66.315000	&	72.805833	&	3.56	\\
UGC04483	&	129.262500	&	69.775278	&	3.58	\\
MESSIER082	&	148.974583	&	69.682500	&	3.61	\\
KDG061	&	149.261250	&	68.591667	&	3.66	\\
NGC2976	&	146.815000	&	67.913611	&	3.66	\\
DDO 053	&	128.527083	&	66.179167	&	3.68	\\
MESSIER081	&	148.889583	&	69.066667	&	3.7	\\
UGC05497	&	153.201667	&	64.107500	&	3.73	\\
F8D1	&	146.196250	&	67.438611	&	3.75	\\
IKN	&	152.024583	&	68.399167	&	3.75	\\
KDG064	&	151.757917	&	67.827500	&	3.75	\\
d0958+66	&	149.702083	&	66.849722	&	3.82	\\
GARLAND	&	150.925000	&	68.693333	&	3.82	\\
d1028+70	&	157.165417	&	70.233611	&	3.84	\\
HolmIX	&	149.385000	&	69.043056	&	3.85	\\
NGC3077	&	150.837500	&	68.733889	&	3.85	\\
KDG073	&	163.237917	&	69.545833	&	3.91	\\
A0952+69	&	149.370833	&	69.272222	&	3.93	\\
DDO 082	&	157.645833	&	70.619444	&	3.93	\\
IC2574	&	157.093333	&	68.416111	&	3.93	\\
HolmI	&	145.134583	&	71.186389	&	4.02	\\
BK3N	&	148.452083	&	68.969167	&	4.17	\\
UGC07298	&	184.119167	&	52.227222	&	4.19	\\
UGC06541	&	173.371250	&	49.238056	&	4.23	\\
d0959+68	&	149.887917	&	68.656944	&	4.27	\\
NGC4068	&	181.010000	&	52.588611	&	4.39	\\
KK109	&	176.796667	&	43.671944	&	4.51	\\
MCG09-20-131	&	183.944583	&	52.387500	&	4.61	\\
UGC06757	&	176.746250	&	61.334722	&	4.61	\\
NGC3738	&	173.952500	&	54.522778	&	5.3	\\
KK135	&	184.894583	&	58.042778	&	5.46	\\
Grapes	&	178.023333	&	54.792222	&	5.96	\\
\hline							
DDO125	&	186.866124	&	43.534035	&	2.61	\\
DDO 99	&	177.733649	&	38.914049	&	2.65	\\
NGC4190	&	183.398633	&	36.632008	&	2.83	\\
NGC4214	&	183.910646	&	36.348362	&	2.88	\\
KDG 90	&	183.714621	&	36.190201	&	2.98	\\
NGC4163	&	183.048224	&	36.219123	&	2.99	\\
MADCASH-2	&	182.474314	&	35.423248	&	3.01	\\
d0934+70	&	143.540382	&	70.150762	&	3.02	\\
Dw0910+7326	&	137.607581	&	73.422892	&	3.21	\\
BK6N	&	158.719439	&	66.011110	&	3.31	\\
MADCASH J07	&	115.584970	&	65.397672	&	3.39	\\
d0926+70	&	141.704262	&	70.493973	&	3.4	\\
d0955+70	&	148.741602	&	70.361245	&	3.45	\\
d1006+67	&	151.778768	&	67.191736	&	3.61	\\
d0939+71	&	144.866137	&	71.337771	&	3.65	\\
KDG063	&	151.226092	&	66.557001	&	3.65	\\
KKH57	&	150.097580	&	63.231898	&	3.68	\\
BK5N	&	151.241237	&	68.254032	&	3.7	\\
d1041+70	&	160.448034	&	70.131159	&	3.7	\\
d1009+68	&	152.221454	&	68.744898	&	3.73	\\
KDG64	&	150.860512	&	68.112680	&	3.75	\\
FM1	&	146.168642	&	68.750778	&	3.78	\\
KK77	&	147.535550	&	67.509495	&	3.8	\\
d0944+69	&	146.074789	&	69.173701	&	3.84	\\
d1014+68	&	153.780248	&	68.726819	&	3.84	\\
dw J0954+6821	&	148.555325	&	68.377482	&	3.87	\\
HS117	&	155.426969	&	71.101060	&	3.96	\\
d1005+68	&	151.241237	&	68.254032	&	3.98	\\
d1015+69	&	153.815748	&	69.009308	&	4.07	\\
J1228+4358	&	187.113727	&	43.999063	&	4.07	\\
NGC4449	&	187.075854	&	44.094677	&	4.27	\\
NGC4244	&	184.353019	&	37.832755	&	4.31	\\
d1006+69	&	151.672831	&	69.949823	&	4.33	\\
KK160	&	190.949440	&	43.695765	&	4.33	\\
NGC4736	&	192.714679	&	41.159063	&	4.41	\\
Cas1	&	31.509985	&	69.038416	&	4.51	\\
IC3687	&	190.587934	&	38.530116	&	4.57	\\
M94-dw2	&	192.701693	&	41.662077	&	4.7	\\
DDO127	&	187.090133	&	37.253574	&	4.72	\\
J1243+41	&	190.979521	&	41.463620	&	4.81	\\
DDO126	&	186.729153	&	37.167847	&	4.97	\\
UGCA281	&	186.604520	&	48.544859	&	5.7	\\
Arp 211	&	189.312032	&	38.782555	&	6.14	\\
NGC4144	&	182.468725	&	46.421076	&	6.89	\\
UGC7639	&	187.472767	&	47.573257	&	7.14	\\
KKH34	&	89.779274	&	73.425345	&	7.28	\\
KDG101	&	184.781960	&	47.042640	&	7.28	\\
DDO120	&	185.332830	&	45.805580	&	7.28	\\
KK132	&	184.760320	&	47.767288	&	7.31	\\
BTS132	&	185.884591	&	47.683344	&	7.4	\\
NGC2787	&	139.820778	&	69.236964	&	7.48	\\
NGC4242	&	184.375712	&	45.597763	&	7.62	\\
NGC4258	&	184.651771	&	47.327853	&	7.66	\\
NGC4618	&	190.396507	&	41.151496	&	7.66	\\
UGC4998	&	141.217257	&	68.391474	&	8.24	\\
\hline
\end{longtable}

\newpage
\section{Propagation of statistical uncertainties}
\label{app:error_propagation}
Given the uncertainties in the measured distances and heliocentric velocities, the propagation of errors is performed as follows:
We correct the heliocentric velocities for the solar motion in the Local Group according to Eq.~\eqref{eq:v_LG},
\begin{align}
v_\mathrm{LG} = v_\mathrm{hel} &+ v_\odot \Big( \sin (b) \sin(b_\odot) + \cos(b) \cos(b_\odot) \cos(l-l_\odot) \Big) \;.
\end{align}
Due to the uncertainties in the apex parameters (Eq.~\ref{eq:apex}), the linear uncertainty propagation
\begin{equation}
\Delta v_\mathrm{LG} = \sqrt{\left(\Delta v_\mathrm{hel}\right)^2 +\left (\frac{\partial v_\mathrm{LG}}{\partial v_\odot}\Delta v_\odot\right)^2 + \left(\frac{\partial v_\mathrm{LG}}{\partial l_\odot}\Delta l_\odot\right)^2 + \left(\frac{\partial v_\mathrm{LG}}{\partial b_\odot}\Delta b_\odot\right)^2} \;,
\end{equation}
with
\begin{align}
\frac{\partial v_\mathrm{LG}}{\partial v_\odot} &= \sin (b) \sin(b_\odot) + \cos(b) \cos(b_\odot) \cos(l-l_\odot) \;, \\
\frac{\partial v_\mathrm{LG}}{\partial l_\odot} &= v_\odot \cos (b) \cos(b_\odot) \sin(l-l_\odot) \;, \\
\frac{\partial v_\mathrm{LG}}{\partial b_\odot} &= v_\odot \left( \sin (b) \cos(b_\odot) - \cos(b) \sin(b_\odot) \cos(l-l_\odot) \right)
\end{align}
is just an approximation to realistic error bars, neglecting uncertainties in $l$ and $b$ as well.

Next, we equip distances to the barycentre with uncertainties. For the simplicity of notation, we define the measured distance to the barycentre and a galaxy~$j$ as $r_\mathrm{c} \equiv \Vert \boldsymbol{r}_\mathrm{c} \Vert_2$ and $r_j \equiv \Vert \boldsymbol{r}_{j} \Vert_2$, respectively. 
The angular distance between the barycentre and galaxy~$j$ is then $\theta_{\mathrm{c}j}$.
In this notation, $r_{\mathrm{c}j} \equiv \Vert \boldsymbol{r}_\mathrm{c} - \boldsymbol{r}_{j}  \Vert_2$, and its uncertainty are given by
\begin{equation}
r_{\mathrm{c}j} = \sqrt{r_\mathrm{c}^2 + r_j^2 - 2r_\mathrm{c} r_j \cos \theta_{\mathrm{c}j}} \;, \quad \Delta r_{\mathrm{c}j} = \sqrt{\left(\frac{r_\mathrm{c}-r_j \cos\theta_{\mathrm{c}j} }{r_{\mathrm{c}j}} \, \Delta r_\mathrm{c} \right)^2 + \left(\frac{r_j-r_\mathrm{c} \cos\theta_{\mathrm{c}j} }{r_{\mathrm{c}j}} \, \Delta r_j\right)^2} \;.
\end{equation}
Uncertainties in $\theta_{\mathrm{c}j}$ are thus considered to be negligible again as in the previous step.
Moreover, if the barycentre is determined as the mass-weighted positional average of several galaxies, the correlation between the uncertainties in $r_\mathrm{c}$ and $r_j$ for those galaxies are neglected as well. 
Since these galaxies are located in the very centre of the M~81 complex and not relevant to constrain the second turnaround radius or the zero-velocity radius, a more precise error estimate is not important for this work.

\newpage
\section{Infall models and barycentre values}
\label{app:individual_infall}

\begin{figure*}[h!]
    \centering
    \includegraphics[width=0.98\linewidth]{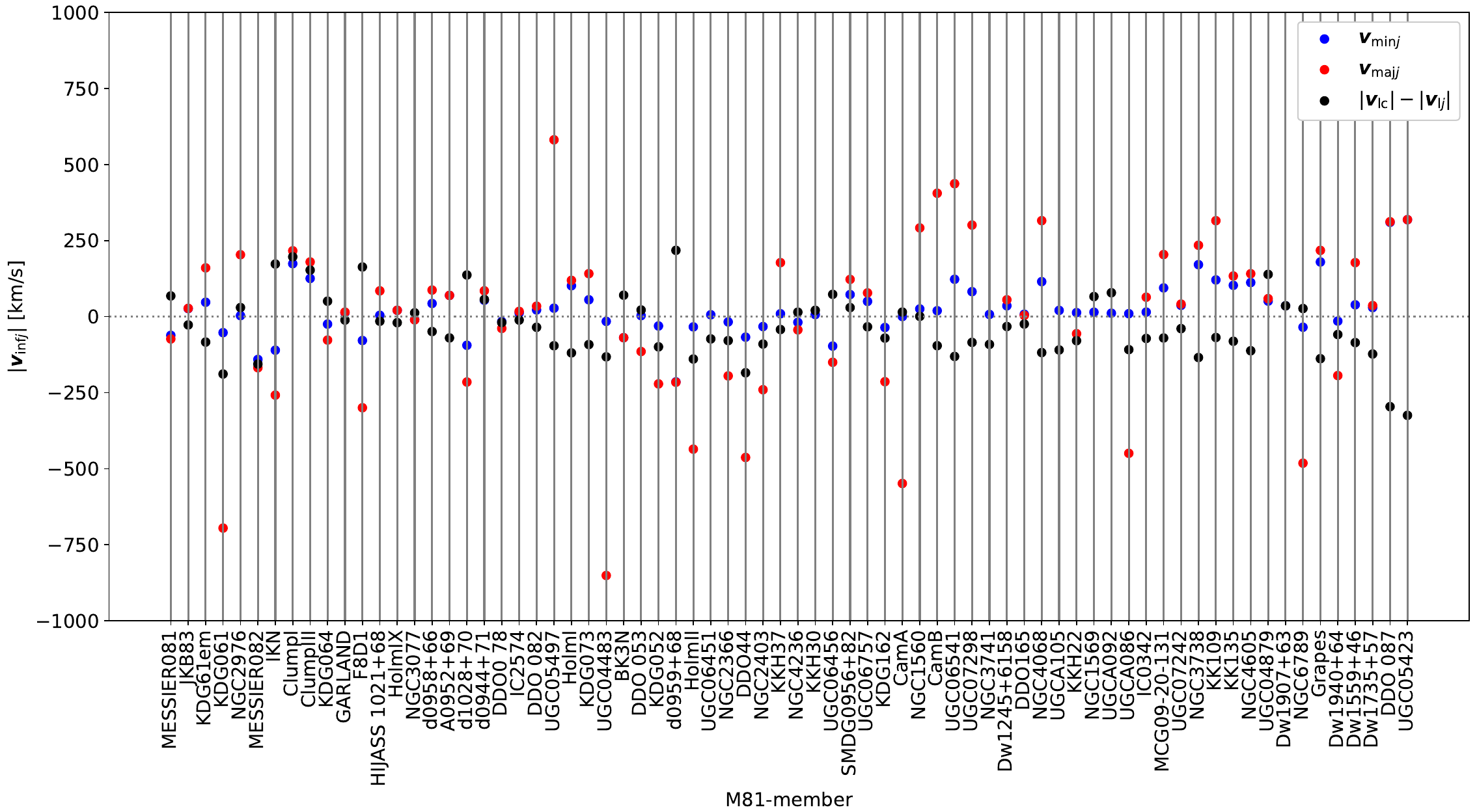}
\caption{\textit{The infall velocities for all 72 galaxies in our data set with respect to the barycentre of the M~81 complex located on the connection line between M~81 and M~82, including velocity corrections for the observers' motion within the Local Group. The galaxies are sorted with increasing distance from M~81 from left to right. The minor infall model velocities are marked in blue, major infall velocities in red, and just taking the difference between the line-of-sight velocities of the barycentre and a galaxy~$j$ are marked in black.}}
    \label{fig:individual_infall}
\end{figure*}

\begin{table*}
\centering
\caption{\it{Hubble flow fit parameters for the minor (top row) and major infall model (centre, bottom row): $\kappa H_0$ in km/s/Mpc (left) and $M$ in $10^{12} M_\odot$ (right) with the singular values as uncertainties. For the major infall, cuts for $\Vert \boldsymbol{v}_{\mathrm{maj}j} \Vert_2 < 700$~km/s were applied (centre row) and $\Vert \boldsymbol{v}_{\mathrm{maj}j} \Vert_2 < 500$~km/s (bottom row). Mass uncertainties were much smaller than the precision for $M$ stated here and are thus omitted.}}

\begin{tabular}{cccc}
\hline
$v$	&	$d_\mathrm{min}$	&		\multicolumn{2}{c}{Barycentre}		\\
	&	$\left[\mbox{Mpc}\right]$	&	M~81	&	M~81 \& M~82	\\
\hline					
\multirow{3}{*}{$v_\mathrm{hel}$}	&	0.5	&	$119.5\pm 27.6$	&	$98.2\pm25.4$	\\
	&	0.75	&	$138.2\pm 27.2$	&	$104.6\pm24.6$	\\
	&	1	&	$157.2\pm 27.0$	&	$117.2\pm24.5$	\\
\hline				
\multirow{3}{*}{$v_\mathrm{LG}$}	&	0.5	&	$104.1\pm 28.0$	&	$96.7\pm 26.6$	\\
    &	0.75	&	$112.3\pm 27.7$	&	$88.1\pm 26.0$	\\
	&	1	&	$120.9\pm 27.5$	&	$95.6\pm25.9$ \\
\hline
\end{tabular}
\begin{tabular}{cccc}
\hline
$v$	&	$d_\mathrm{min}$	&		\multicolumn{2}{c}{Barycentre}		\\
	&	$\left[\mbox{Mpc}\right]$	&	M~81	&	M~81 \& M~82	\\
\hline					
\multirow{3}{*}{$v_\mathrm{hel}$}	&	0.5	&	$12.5$	&	$7.6$	\\
	&	0.75	&	$20.8$	&	$9.9$	\\
	&	1	&	$31.7$	&	$15.5$	\\
\hline							
\multirow{3}{*}{$v_\mathrm{LG}$}	&	0.5	&	$4.8$	&	$4.1$	\\
    &	0.75	&	$7.0$	&	$2.8$	\\
	&	1	&	$9.7$	&	$4.5$ \\
\hline
\end{tabular}
\\[2ex]
\begin{tabular}{cccc}
\hline
$v$	&	$d_\mathrm{min}$	&		\multicolumn{2}{c}{Barycentre}		\\
	&	$\left[\mbox{Mpc}\right]$	&	M~81	&	M~81 \& M~82	\\
\hline					
\multirow{3}{*}{$v_\mathrm{hel}$}	&	0.5	&	$327.3\pm 41.7$	&	$361.2\pm 39.9$	\\
	&	0.75	&	$466.1\pm 39.9$	&	$661.0\pm 38.6$	\\
	&	1	&	$2525.9\pm 37.1$	&	$2178.0\pm 37.1$	\\
\hline							
\multirow{3}{*}{$v_\mathrm{LG}$}	&	0.5	&	$303.2\pm 40.8$	&	$167.9\pm 35.2$	\\
    &	0.75	&	$440.2\pm 38.5$	&	$205.7\pm 32.6$	\\
	&	1	&	$-502.8\pm 35.1$	&	$280.9\pm 27.1$ \\
\hline
\end{tabular}
\begin{tabular}{cccc}
\hline
$v$	&	$d_\mathrm{min}$	&		\multicolumn{2}{c}{Barycentre}		\\
	&	$\left[\mbox{Mpc}\right]$	&	M~81	&	M~81 \& M~82	\\
\hline					
\multirow{3}{*}{$v_\mathrm{hel}$}	&	0.5	&	$172.6$	&	$228.2$	\\
	&	0.75	&	$434.3$	&	$946.1$	\\
	&	1	&	$18570.0$	&	$13078.7$	\\
\hline							
\multirow{3}{*}{$v_\mathrm{LG}$}	&	0.5	&	$132.4$	&	$39.2$	\\
    &	0.75	&	$363.0$	&	$71.1$	\\
	&	1	&	$1098.5$	&	$165.7$ \\
\hline
\end{tabular}
\\[2ex]
\begin{tabular}{cccc}
\hline
$v$	&	$d_\mathrm{min}$	&		\multicolumn{2}{c}{Barycentre}		\\
	&	$\left[\mbox{Mpc}\right]$	&	M~81	&	M~81 \& M~82	\\
\hline					
\multirow{3}{*}{$v_\mathrm{hel}$}	&	0.5	&	$312.1\pm 37.6$	&	$341.5\pm 38.6$	\\
	&	0.75	&	$421.3\pm 36.9$	&	$706.6\pm 37.3$	\\
	&	1	&	$924.9\pm 35.4$	&	$-5740.4\pm 35.5$	\\
\hline							
\multirow{3}{*}{$v_\mathrm{LG}$}	&	0.5	&	$339.5\pm 37.6$	&	$156.4\pm 33.6$	\\
    &	0.75	&	$572.6\pm 36.8$	&	$192.4\pm 30.6$	\\
	&	1	&	$-502.8\pm 35.1$	&	$280.9\pm 27.1$ \\
\hline
\end{tabular}
\begin{tabular}{cccc}
\hline
$v$	&	$d_\mathrm{min}$	&		\multicolumn{2}{c}{Barycentre}		\\
	&	$\left[\mbox{Mpc}\right]$	&	M~81	&	M~81 \& M~82	\\
\hline					
\multirow{3}{*}{$v_\mathrm{hel}$}	&	0.5	&	$170.2$	&	$197.5$	\\
	&	0.75	&	$374.7$	&	$1099.8$	\\
	&	1	&	$2358.3$	&	$100123.3$	\\
\hline							
\multirow{3}{*}{$v_\mathrm{LG}$}	&	0.5	&	$180.2$	&	$31.4$	\\
    &	0.75	&	$691.5$	&	$58.9$	\\
	&	1	&	$1098.5$	&	$165.7$ \\
\hline
\end{tabular}

\label{tab:HFF_svd_results}
\end{table*}



\end{document}